\documentclass[usenatbib]{mnras}
\usepackage{newtxtext,newtxmath}
\usepackage[T1]{fontenc}
\DeclareRobustCommand{\VAN}[3]{#2}
\let\VANthebibliography\thebibliography
\def\thebibliography{\DeclareRobustCommand{\VAN}[3]{##3}\VANthebibliography}
\newcommand{\myref}[1]{Eq.\,\eqref{#1}}

\usepackage{graphicx}	% Including figure files
\usepackage{amsmath}	% Advanced maths commands

\usepackage{amssymb}	% Extra maths symbols
\usepackage{multirow}
\usepackage{indentfirst}
\usepackage{url}
\usepackage{gensymb}

\newcommand{\ba}{\begin{eqnarray}}
\newcommand{\ea}{\end{eqnarray}}

\title[DESI groups $\times$ Planck CMB lensing]{Cross-correlation of \emph{Planck} CMB lensing with DESI galaxy groups}

\author[Sun et al.]{
Zeyang Sun,$^{1,2}$\thanks{E-mail: zeyangsun@sjtu.edu.cn}
Ji Yao,$^{1,2}$
Fuyu Dong,$^{3}$
Xiaohu Yang,$^{1,2,4}$
Le Zhang,$^{5}$
Pengjie Zhang$^{1,2,4}$\thanks{E-mail: zhangpj@sjtu.edu.cn}
\\
$^{1}$Department of Astronomy, School of Physics and Astronomy, Shanghai Jiao Tong University, Shanghai, 200240, China\\
$^{2}$Shanghai Key Laboratory for Particle Physics and Cosmology, Shanghai, 200240, China\\
$^{3}$School of Physics, Korea Institute for Advanced Study (KIAS), 85 Hoegiro, Dongdaemun-gu, Seoul, 02455, Republic of Korea\\
$^{4}$Division of Astronomy and Astrophysics, Tsung-Dao Lee Institute, Shanghai Jiao Tong University, Shanghai, 200240, China\\
$^{5}$School of Physics and Astronomy, Sun Yat-Sen University, 2 Daxue Road, Tangjia, Zhuhai, 519082, China
}

\date{Accepted XXX. Received YYY; in original form ZZZ}

\pubyear{2021}

\begin{document}
\label{firstpage}
\pagerange{\pageref{firstpage}--\pageref{lastpage}}
\maketitle

\begin{abstract}
We measure the cross-correlation between galaxy groups constructed from DESI Legacy Imaging Survey DR8 and \emph{Planck} CMB lensing, over overlapping sky area of 16876 $\rm deg^2$. The detections are significant and consistent with the expected signal of the large-scale structure of the universe, over group samples of various redshift, mass, richness $N_{\rm g}$ and over various scale cuts. The overall S/N is 40 for a conservative sample with $N_{\rm g}\geq 5$, and increases to $50$ for the sample with $N_{\rm g}\geq 2$. Adopting the \emph{Planck} 2018 cosmology, we constrain the density bias of groups with $N_{\rm g}\geq 5$ as $b_{\rm g}=1.31\pm 0.10$, $2.22\pm 0.10$, $3.52\pm 0.20$ at $0.1<z\leq 0.33$, $0.33<z\leq 0.67$, $0.67<z\leq1$ respectively. The group catalog provides the estimation of group halo mass and therefore allows us to detect the dependence of bias on group mass with high significance. It also allows us to compare the measured bias with the theoretically predicted one using the estimated group mass. We find excellent agreement for the two high redshift bins. However, it is lower than the theory by $\sim 3\sigma$ for the lowest redshift bin. Another interesting finding is the significant impact of the thermal Sunyaev Zel'dovich (tSZ). It contaminates the galaxy group-CMB lensing cross-correlation at $\sim 30\%$ level, and must be deprojected first in CMB lensing reconstruction. 
\end{abstract}


%%%%%%%%%%%%%%%%%%%%%%%%%%%%%%%%%%%%%%%%%%%%%%%%%%

%%%%%%%%%%%%%%%%% BODY OF PAPER %%%%%%%%%%%%%%%%%%
\section{Introduction}
There is a wealth of cosmological information in secondary anisotropies resulting from perturbations to Cosmic Microwave Background (CMB) light after the time of recombination. A particularly interesting source of secondary anisotropy is the gravitational lensing \citep{Seljak1996, Seljak1999, Hu2000, Hu2002, Okamoto2003}, which deflects  the paths of photons from the last-scattering surface. These deflections, on the order of a few arc-minutes, alter the CMB primary anisotropies by re-distributing power across different angular scales and producing a non-Gaussian and anisotropic component to the primordial distribution of temperature anisotropies. Through such  non-Gaussian and anisotropic features, the gravitational field can be reconstructed from the observed CMB maps \citep{Seljak1999, Hu2002, Okamoto2003, Lewis2006}. The first detection of CMB lensing is by the Atacama Cosmology Telescope (ACT) \citep{ACT2011}. High signal-to-noise measurements of CMB lensing have been obtained by several collaborations, including a full-sky map of lensing convergence by the \emph{Planck} satellite mission \citep{Planck2018parameters, Planck2020c}, a 2500 $\rm deg^2$ CMB lensing map from combined South Pole Telescope (SPT) and \emph{Planck} temperature data \citep{SPT2017}, a measurement of the lensing B-mode CMB polarization by POLARBEAR \citep{POLARBEAR2014}, and most recently a lensing map over 2100 $\rm deg^2$  by ACT \citep{ACT2021}.

The CMB lensing signal is an integral of all deflections sourced by the large-scale structure (LSS) between the last-scattering surface and the observer, thus it contains rich cosmological information with its auto-correlation. Meanwhile, it is physically correlated with other LSS tracers such as galaxy and galaxy cluster distribution, and cosmic shear. On one hand, these cross-correlations are immune to certain systematics in the auto-correlation, such as the additive errors in both CMB lensing and cosmic shear. On the other hand, they provide essential information, such as the redshift information, to improve the cosmological applications of CMB lensing. CMB lensing-cosmic shear cross-correlation has been detected by various data sets \citep{Hand2015, Liu2015, Kirk2016, Harnois2016, Singh2017, Harnois2017, Omori2019b, Namikawa2019, Marques2020, Robertson2021}. Cross-correlations of CMB lensing with galaxies (e.g. recent works by \citet{Singh2017, Schmittfull2018, Bianchini2018, Baxter2018, Raghunathan2018, Giusarma2018, Omori2019a, Omori2019b, unWISE2020, Marques2020galaxyclustering, ACT2021, MartinWhite_LRGs, Hang2021, ShekharSaraf2021, Fang2021, Krolewski2021, Alonso2021, Yan2021, White2021}), galaxy groups and clusters \citep{Madhavacheril2015, Baxter2015, Geach2017, Baxter2018, Madhavacheril2020}, and quasars \citep{Lin2020} have also been detected. Recently, the cross-correlation between CMB lensing and a new LSS tracer (low-density positions (LDPs) \citep{Dong2019}, has been detected with S/N $\sim 55 $ \citep{Dong2021}. 

Recently \cite{Yang2020} (hereafter Y21) identified $\sim 10^7$ galaxy groups from the DESI imaging surveys \citep{Dey2019}, over $\sim 20,000$ deg$^2$ and $z<1$. Compared with the majority of galaxies in the DESI imaging surveys, these galaxy groups have better photo-$z$ information and less imaging systematics, making them attractive in CMB lensing-LSS cross-correlation measurement. Furthermore, Y21 also provides an estimation of the halo mass. Such information is crucial in validating/interpreting the cross-correlation measurement. For these reasons, we measure the cross-correlation between the galaxy group positions (number density) of the Y21 catalog and \emph{Planck} CMB lensing \citep{Planck2020c}. With $\sim$40\% overlapping sky coverage and $2 \times 10^6$ galaxy groups of richness $N_{\rm g}\geq 5$ over $0.1<z<1.0$, we detect the cross-correlation signal with S/N $\simeq$ 40. The S/N increases to $50$ if we relax the richness cut to $N_{\rm g}\geq 2$. Thanks to a large number of clusters and groups, we have improved the S/N of cluster/group-CMB cross-correlation measurement by a factor $\ga 5$, over previous measurements \citep{Madhavacheril2015, Baxter2015, Geach2017, Baxter2018, Madhavacheril2020}. An important improvement to previous works is that, we are now able to directly compare the measured group bias $b_{\rm g}$ with the theoretically predicted one using the estimated halo mass of each group. The agreement is excellent for the two higher redshift bins. 

This paper is structured as follows. We first review in Sec. \ref{section:theory} the theoretical foundations of CMB lensing and galaxy clustering; we then describe the DESI imaging DR8 group data and \emph{Planck} legacy data we used in Sec. \ref{section:data}, and the analysis methods we follow in Sec. \ref{section:methods}. The main results of this paper, together with their cosmological implications, are presented in Sec. \ref{section:results}. We finally conclude in Sec. \ref{section:conclusions}.

%%%%%%%%%%%%%%%%% BODY OF PAPER %%%%%%%%%%%%%%%%%%
\section{Theoretical background}\label{section:theory}
Galaxy/galaxy group overdensity $\delta_{\rm g}$ and CMB lensing convergence $\kappa$ are both projections of 3D density fields, expressed as line-of-sight integrals over their respective projection kernels. The angular cross-correlation power spectrum, adopting the Limber approximation \citep{Limber1953}, is 
\begin{equation}
    C_\ell^{\kappa g} =\int d\chi W^\kappa(\chi)W^g(\chi)\frac{1}{\chi^2}P_{mg}\left(k = \frac{\ell+1/2}{\chi};z\right) \ .
\label{eqn:ckg}
\end{equation}

The Limber approximation is inaccurate for $\ell<10$, but such very large-scale modes are excluded from our fitting anyway, due to poor S/N. The above expression assumes spatial flatness.  Here $W^\kappa$ and $W^g$ are the projection kernels for $\kappa$ and the group number density fields.

\begin{equation}\label{eq:Wg}
    W^g(z) = n(z) = \frac{c}{H(z)}W^g(\chi) \ .
\end{equation}
\begin{equation}\label{eq:Wk}
    W^\kappa(z) = \frac{3}{2c}\Omega_{m0}\frac{H_0^2}{H(z)}(1+z)\frac{\chi(\chi_*-\chi)}{\chi_*} = \frac{c}{H(z)}W^\kappa(\chi) \ .
\end{equation}

Here $\chi$ is the comoving distance to redshift $z$ and $\chi_* = \chi(z_*\approx1089)$ is the distance to the surface of the last scattering. $n(z)$ is the normalized redshift distribution of galaxy groups. These kernels are plotted in Fig.\,\ref{fig:nz}. 
$P_{\rm mg}$ is the 3D cross power spectrum between matter and group number overdensity.  We define the group bias through $b_{\rm g}\equiv P_{\rm mg}/P_{\rm mm}$. $b_{\rm g}$ is approximately scale-independent at the large scales of interest. 

\begin{figure}
    \centering
    \includegraphics[width=\columnwidth]{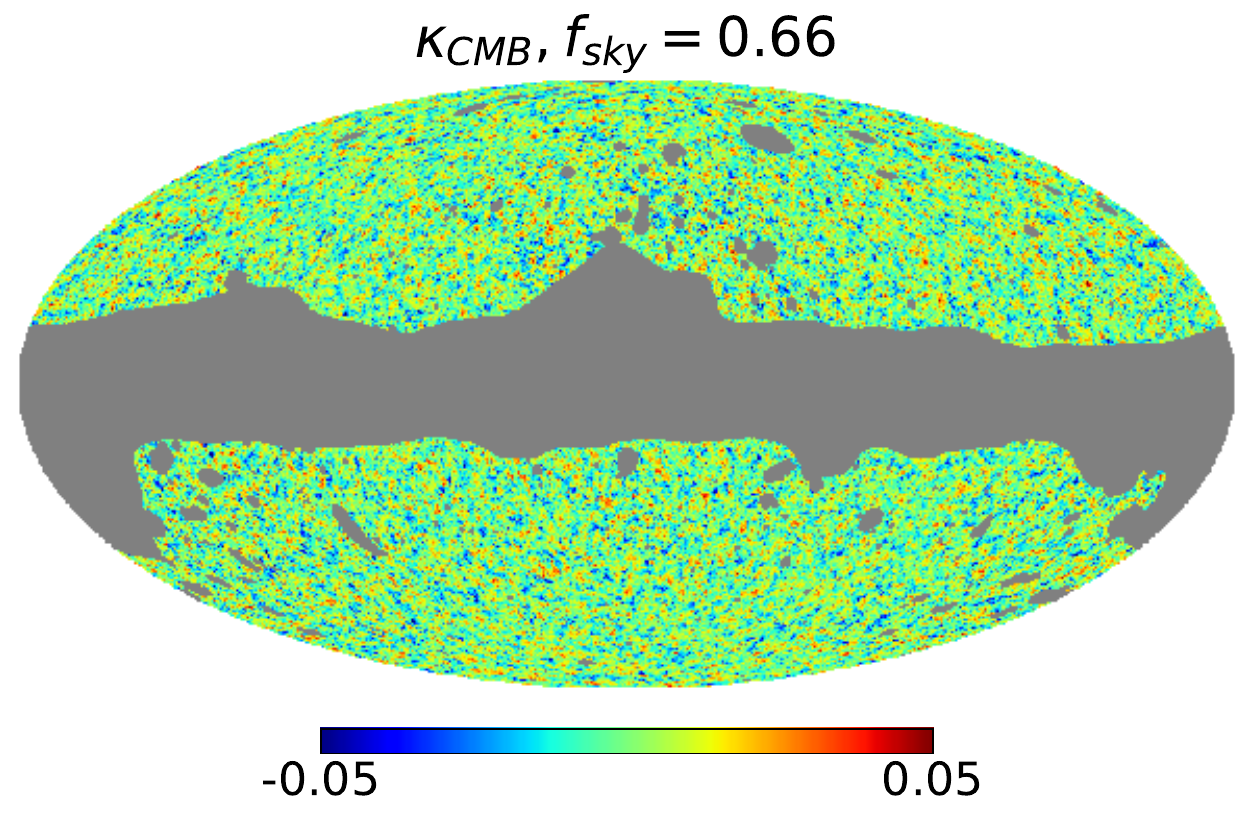}
    \caption{\emph{Planck} 2018 CMB lensing map filtering by Wiener filter. The map is constructed in HEALPix pixelization scheme, with resolution parameter $N_{\rm side} = 512$ (6.9 arcmin). The mask (grey area) is \emph{Planck} tSZ-deprojected CMB lensing mask.}
    \label{fig:kappa_ma}
\end{figure}

\section{DATA}\label{section:data}
\subsection{\emph{Planck} CMB lensing maps}
We use the lensing convergence maps and analysis masks provided in the \emph{Planck} 2018 release\footnote{\url{http://pla.esac.esa.int/pla/\#cosmology}} \citep{Planck2020c}. To minimize the impact of tSZ, we adopt the baseline lensing map as the one reconstructed from the tSZ-deprojected temperature-only SMICA map. This map has a lower S/N than the lensing map reconstructed combining temperature and polarization. We also repeat the analysis for this map. Hereafter the two maps will be referred to as {\bf BASE} and {\bf with tSZ}, respectively.

The spherical harmonics coefficients of lensing convergence are provided in HEALPix \citep{Zonca2019, Gorski2005} format with maximum order $\ell_{\rm max}$ = 4096, and the associated analysis masks are given as HEALPix maps with resolution $N_{\rm side}$ = 2048 \citep{Planck2020c}. We are only interested in large scale clustering and we prefer to avoid small scale complexities such as scale-dependent bias, baryonic effect, and massive neutrino \citep{DESY3, Heymans2021, Asgari2021}. Therefore we downgrade the mask to $N_{\rm side}$ = 512, with the lensing harmonics up to $\ell_{\rm max}$ = 1000. The downgraded \emph{Planck} lensing map is shown in Fig.\,\ref{fig:kappa_ma}. Notice that direct downgrading may result in a large aliasing effect, due to overwhelming reconstruction noise at pixelization scale. Thus we first apply a Wiener filter provided by the \emph{Planck} 2018 release to the original convergence map, then downgrade it to $N_{\rm side}$ = 512.

\subsection{DESI group samples}
\begin{figure}
    \centering
    \includegraphics[width=\columnwidth]{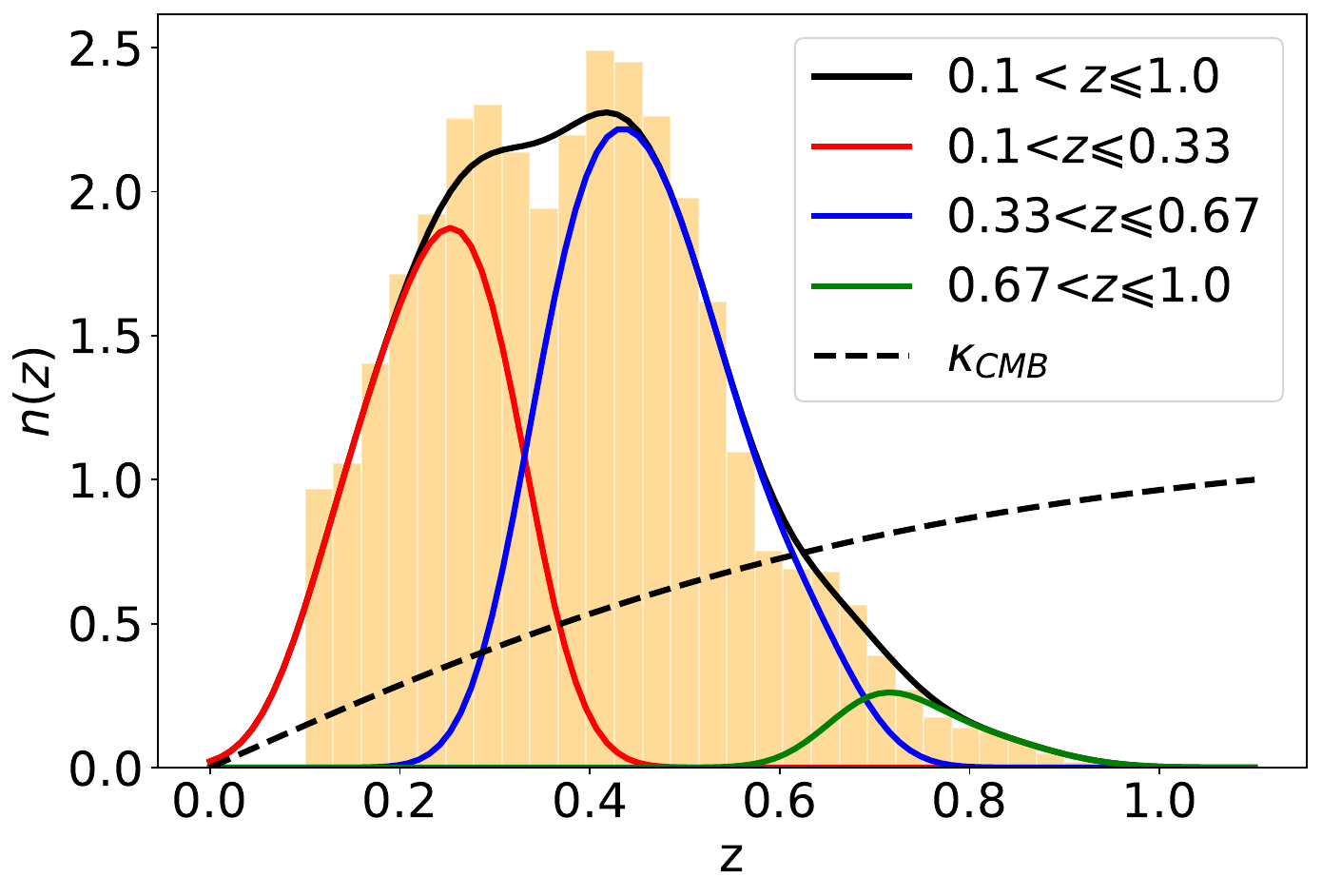} 
    \caption{Redshift distribution with three tomographic bins of galaxy group samples in the photo-$z$ range (0.1,1.0]. The histogram is the photo-$z$ distribution and the corresponding color curve is the estimated true redshift distribution (normalized) with photo-$z$ error $\sigma_{z}=0.05$. Projection kernels for the group sample (solid curves) and CMB lensing (dashed black curve), expressed in \myref{eq:Wg} and \myref{eq:Wk} respectively, both normalized to unit maximum.}
    \label{fig:nz}
\end{figure}

\begin{figure*}
    \centering
    \includegraphics[scale=0.35]{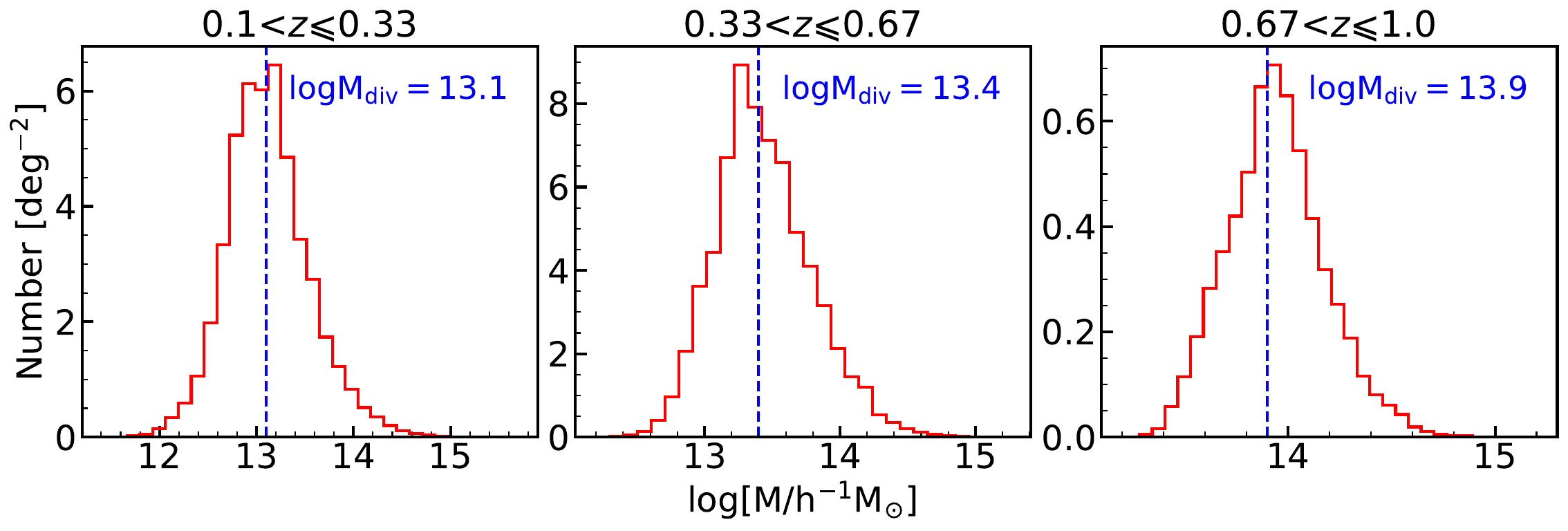} 
    \caption{Mass distribution of galaxy groups in three tomographic bins within photo-$z$ range (0.1,1.0]. The value $\log M_{\rm div}$ is defined as the mass of dividing the groups into two mass bins, low-mass and high-mass, that have the same quantity of groups.}
    \label{fig:mass_distribution}
\end{figure*}

\begin{table}
    \centering
    \begin{tabular}{cccccccc} 
    	\hline
    	$z$ &  of groups &  with $N_{g}\geqslant5$ & $\log M_{\rm div}$ \\
    	\hline 
    	$0.1<z\leqslant0.33$ &20151147 &800606& 13.1 \\
        $0.33<z\leqslant0.67$ & 49901082 &1134482& 13.4\\
        $0.67<z\leqslant1.0$ & 20829772 &101676& 13.9\\
    	\hline
    \end{tabular}
    \caption{The information of groups in the DESI Imaging Survey DR8. Column (2) lists the number of groups in the original catalog. Column (3) lists the number of groups with at least 5 members. Column (4) is the mass we divide the samples into two mass bins (low and high) in each redshift bin, and both mass bins have equal size of groups. }
    \label{tab:list}
\end{table}

The Y21 catalog is constructed from the DESI Legacy Imaging Survey DR8, with a halo-based group finder. Each group is assigned an estimated total halo mass (estimated based on abundance matching), \{ra, dec\} of the brightest galaxy, photometric redshift, richness $N_{\rm g}$ (the number of member galaxies in each group), etc \citep{Yang2020}. Fig.\,\ref{fig:nz} shows the redshift distribution of groups with $N_{\rm g}\geq 5$, in the three tomographic bins ($0.1<z\leq 0.33$, $0.33<z\leq 0.67$, $0.67<z<1$). Table. \ref{tab:list} lists the total number of groups in each redshift bin. The first two redshift bins have $0.8$M and $1.1$M groups respectively. The last redshift bin has only $0.1$M groups. But these groups have stronger clustering and redshifts closer to the peak of the lensing kernel. Therefore they produce significant cross-correlation S/N. We also divide groups in each redshift bin into two mass bins of an equal amount of groups. Fig.\,\ref{fig:mass_distribution} show the mass distribution of groups with three redshift bins. $\log M_{\rm div}$ is defined as the mass of dividing the groups into two mass bins that have an equal quantity of groups. As expected, $\log M_{\rm div}$ increases with redshift, what we could observe in higher redshift are more massive galaxy groups.

\subsection{Survey mask}\label{subsec:survey mask}
As described in Y21, the galaxy mask is constructed from a set of uniformly distributed random catalogs provided in the DESI DR8 website \footnote{\url{http://legacysurvey.org/dr8/files/\#random-catalogs}}. We adopt the identical group mask as in \cite{Yang2020}. 

Each random point contains the exposure times for $g,r,z$ bands based upon the sky coordinate drawn independently from the observed distribution. We choose random points whose exposure time in all three bands is greater than zero to produce the survey masks which populate the same sky coverage and geometry with the galaxy catalog. In addition to the Tractor morphology \footnote{‘Tractor’ is the fitting procedure that DESI used officially. The Tractor code runs within the geometrical region of a brick to produce catalogs of extracted sources. In DR8, six morphological types are used. Five of these are used in the Tractor fitting procedure: point sources, round exponential galaxies with a variable radius ("REX"), deVaucouleurs ("DEV") profiles (elliptical galaxies), exponential ("EXP") profiles (spiral galaxies), and composite profiles that are deVaucouleurs + exponential (with the same source center). The sixth morphological type is "DUP," which is set for Gaia sources that are coincident with, and so have been fit by, an extended source. Website: \url{https://legacysurvey.org/dr8/description/tractor-catalogs}} selection, minimum data quality flags are employed to remove the flux contaminations from nearby sources (FRACFLUX) or masked pixels (FRACMASKED):
\begin{center}
    FRACMASKED\_i < 0.4 \\
    FRACIN\_i > 0.3 \\
    FRACFLUX\_i < 0.5
\end{center}
where i $\equiv g,r,z$. FRACIN is to select the objects for which a large fraction of the model flux is in the $N_{\rm side}$ = 512 contiguous pixels where the model was fitted. We also remove any objects close to the bright stars by masking out objects with the following bits number in the DR8 MASKBITS column: 1 (close to Tycho-2 and GAIA bright stars), 8 (close to WISE W1 bright stars), 9 (close to WISE W2 bright stars), 11 (close to fainter GAIA stars), 12, and 13 (close to an LSIGA large galaxy and a globular cluster, respectively). 

In addition, we project the 3D group number density distribution to 2D sky map along the line-of-sight in the $N_{\rm side}$ = 2048 resolution, then downgrade the pixelized map to $N_{\rm side}$ = 512, and exclude areas pixels with an observed coverage fraction smaller than $f^{\rm threshold}$. This step is to minimize the impact from the footprint-induced fake overdensity. For the pixels with coverage fraction above this threshold, we assign to each mask pixel $\alpha$ its coverage fraction $f_\alpha$, and use this value as a weight in the clustering measurements that follow
\begin{equation}
    f_{\alpha} = \frac{\sum\limits_{i=1}^{16}w_{i}}{\sum\limits_{i}} \geqslant f^{\rm threshold} = 0.5 \label{eq:fi} \ ,
\end{equation}
where $w_i$ is the galaxy mask in $N_{\rm side}=2048$ map, which is 1 when there are galaxies satisfying group selection criteria otherwise is 0. And the denominator in \myref{eq:fi} is 16. We consider three choices of $f^{\rm threshold}$ later for consistency tests but fiducial one is 0.5, and the group mask is 1 (meaning the pixel will be used in the cosmological analysis) if $f_\alpha\geqslant f^{\rm threshold}$ otherwise is 0 (excluded in the analysis). 

% \begin{figure*}
%     \centering 
%     \includegraphics[scale=0.35]{figures/fig3.pdf}
%     \caption{\color{red}{The number overdensity maps of galaxy groups at three redshift bins and two mass bins. The maps are constructed in HEALPix pixelization scheme, with resolution parameter $N_{\rm side} = 512$, corresponding to to 6.9 arcmin. The detailed description of group mask (grey area) is in \S \ref{subsec:survey mask}, and the coverage fraction of sky is 0.43.}}
%     \label{fig:ModifiedMap}
% \end{figure*}

The pixelized overdensity of galaxy groups is estimated by
\begin{equation}
    \hat{\delta}_\alpha = \frac{N_\alpha / f_\alpha}{\langle N_\alpha / f_\alpha\rangle} -1 \label{eq:overdensity} \ .
\end{equation}
Here $N_\alpha$ is the number of groups in the $\alpha$-th pixel. The average $\langle \cdots \rangle$ is over all pixels with $f\geq f^{\rm threshold}$. The number overdensity maps are constructed in HEALPix pixelization scheme with resolution parameter $N_{\rm side} = 512$ (6.9 arcmin), and the coverage fraction of groups is 0.43.

The total mask that we use for the cross-correlation analysis is defined as the modified group mask in combination with the \emph{Planck} convergence map mask. After the total masking, there are $16876$ deg$^2$ sky area ($f_{\rm sky}=0.41$) and 2 million galaxy groups left. The reason we choose the combined mask is, we count the pixel pairs between the galaxy group overdensity map and CMB lensing convergence map. The combined mask can correctly count for the number of usable pairs.

\section{Analysis Methods}\label{section:methods}
In this section, we discuss our method of estimating the cross power spectra, as well as their covariance matrix. We also show the theoretical calculation process for cross spectra and the halo bias constraints with the associated halo mass uncertainty. In this paper, we measure the galaxy groups cross-correlate with the CMB lensing maps in harmonic space.

\subsection{Cross-correlation measurements}
Starting with the above maps, we use HEALPix to measure the spherical harmonic coefficients $\delta_{\ell m}$ (group overdensity) and $\kappa_{\ell m}$. The observed cross-power spectrum is then
\begin{equation} 
\label{Eq Cell sum}
    C_\ell^{\kappa g} = \frac{1}{2\ell+1}\sum\limits_{m=-\ell}^\ell \delta_{\ell m}\kappa_{\ell m}^*  \ .
\end{equation}
Given the existence of the survey mask, the raw power spectrum obtained above is biased. This problem can be corrected by deconvolving the survey mask, or by measuring the correlation function instead. Alternatively, the survey mask can be included in the theory side. We take this later approach, detailed in \S \ref{sec 4.2 theory Cl}.

\begin{figure*}
    \centering
    \includegraphics[scale=0.35]{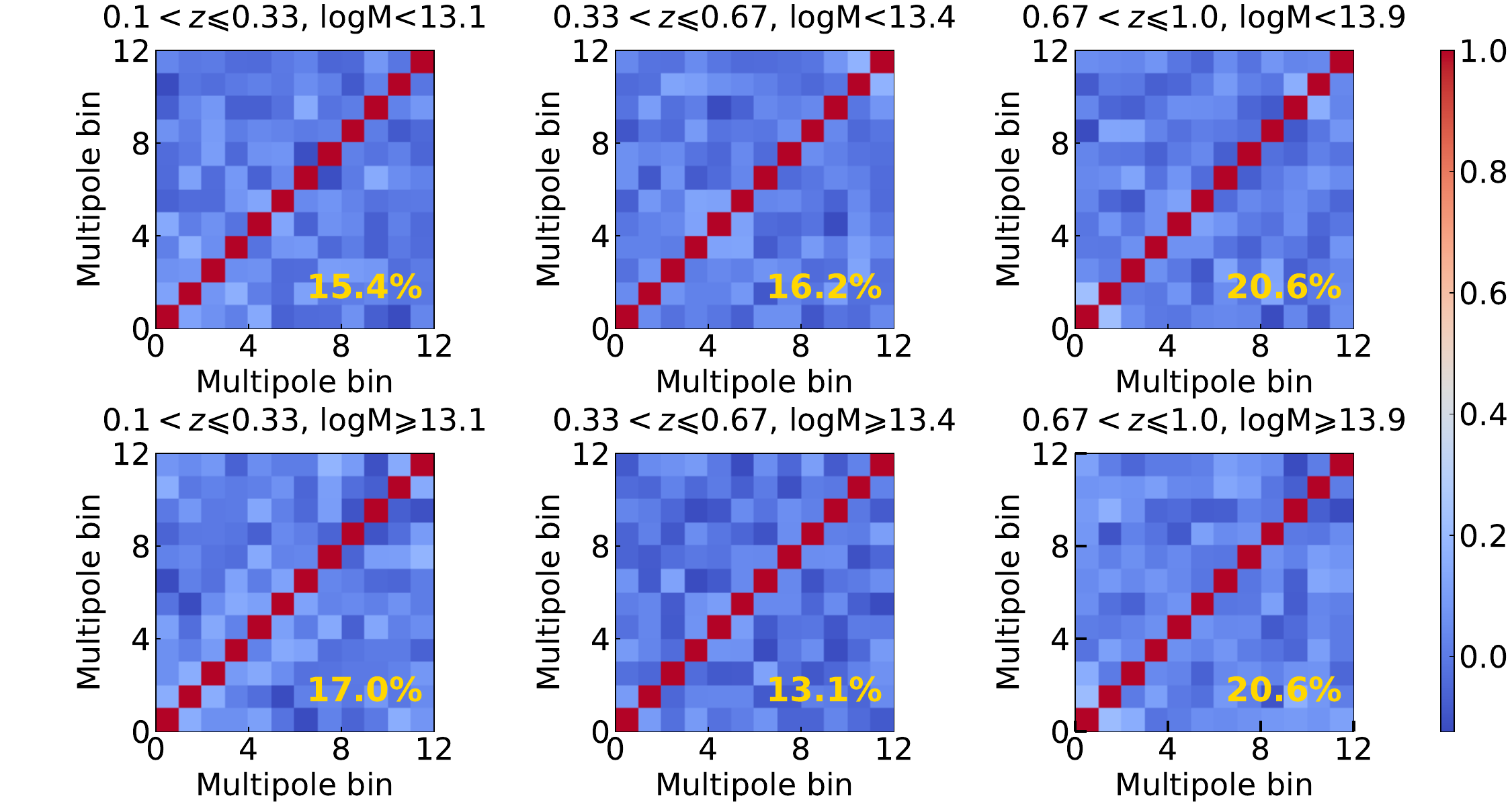}
    \caption{The normalized covariance matrix $C_{ij}/\sqrt{C_{ii}C_{jj}}$ for $C_{\ell}^{\kappa g}$. It is calculated with 300 \emph{Planck} CMB lensing simulation maps by \myref{eq:covariance}. The percentage refers to ratio of the maximum absolute value on the off-diagonal elements to the diagonal elements.}
    \label{fig:r}
\end{figure*}

\begin{figure}
    \centering
    \includegraphics[width=\columnwidth]{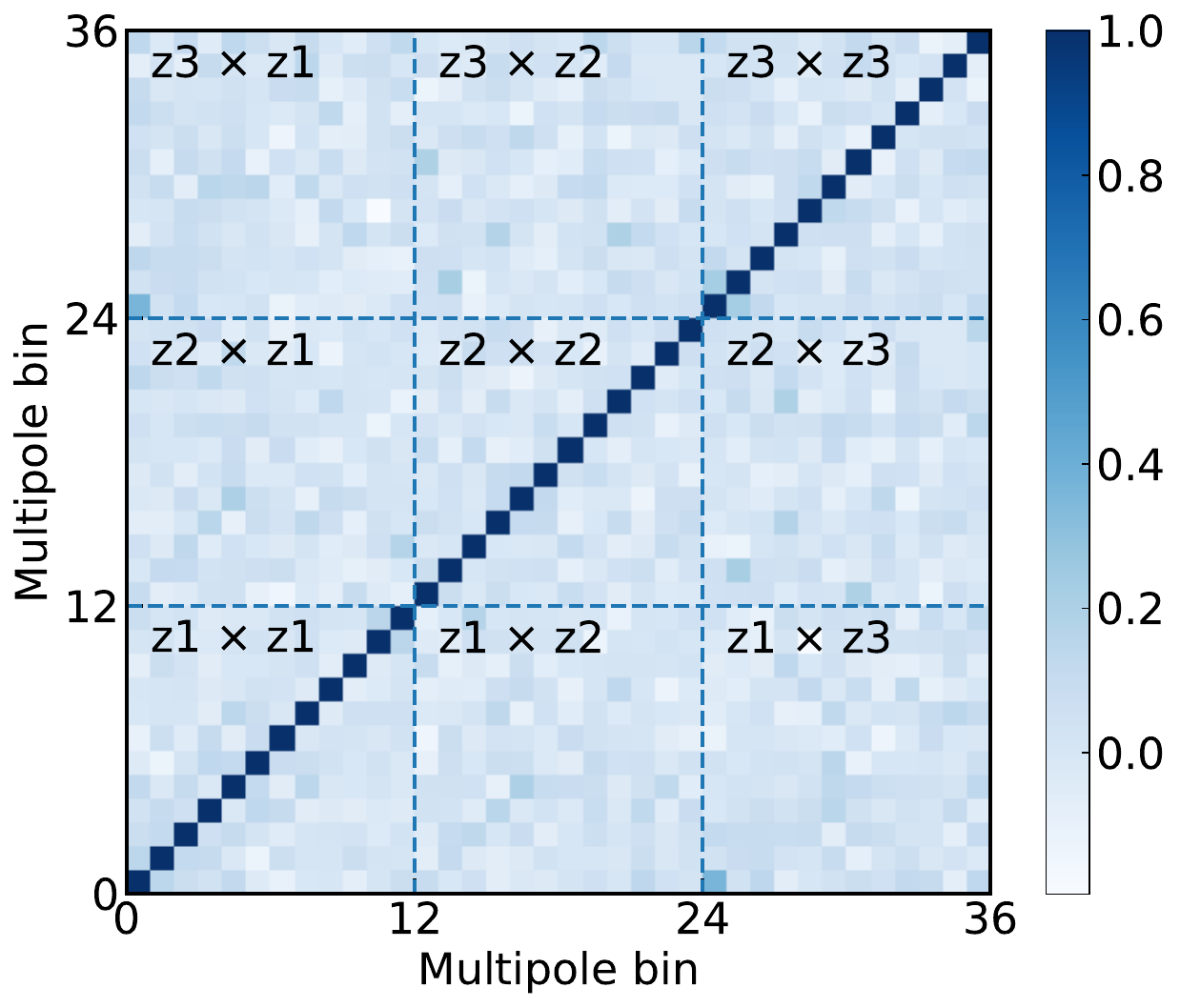}
    \caption{The correlation matrix for cross power spectra at three redshift intervals. Here multipole 0-12 denote bins in the spectrum of the first redshift interval and 12-24, 24-36 denote those of second and third redshift intervals.}
    \label{fig:cov_z123}
\end{figure}

The lensing map is reconstruction noise dominated. The group distribution-CMB lensing cross-correlation coefficient is much smaller than unity, due to mismatch in their redshift distribution. For the two reasons, cosmic variance arising from the group-CMB lensing cross-correlation is negligible in the covariance matrix. Therefore we estimate the covariance using 300 \emph{Planck} CMB lensing simulations \citep{Planck2020c}.  We obtain the mean-field $\langle a_{\ell m}\rangle$ by using first 60 simulations and then subtract them from the remaining 240 simulations to obtain mean-field subtracted maps\footnote{Because \emph{Planck} website offers CMB lensing simulations without mean-field subtraction, but observation data is mean-field subtracted.}. We measure the cross-power spectra of these 240 maps with the overdensity maps, and estimate the covariance matrix by
\begin{equation}
\label{eq:covariance}
    \textbf{Cov}_{\ell\ell^{\prime}} = \frac {1}{N-1} \sum\limits_{n=1}^{N=240} [(C_{n}^{\kappa g}(\ell) - \overline{C}^{\kappa g}(\ell))\times (C_{n}^{\kappa g}(\ell^{\prime}) - \overline{C}^{\kappa g}(\ell^{\prime})) ] \ .
\end{equation}
Here $\overline{C}^{\kappa g}(\ell)$ is the average cross-power spectrum. Fig.\,\ref{fig:r} shows the normalized covariance matrix in each mass bin and tomographic bin. We note the covariance will require correction due to the limited number of realizations and error propagation to the fitting parameters \citep{Hartlap2007}. We test Eq. 32 and 33 in \cite{Wang2020}. The correction factors are $\sim$ 0.95 and $\sim$ 1.03. As a result, the total correction is $\sim$ 1.07 to the covariance, meaning our errors of power spectra are underestimated by about 4\%, reducing our S/N by $\sim$ 1. Fig.\,\ref{fig:cov_z123} shows the correlation matrix for cross power spectra at $0.1<z\leq 0.33$, $0.33<z\leq 0.67$, $0.67<z\leq1$, respectively. As one can see, the off-diagonal blocks can be negligible. That means we could sum the S/N in each redshift bin as defined in \myref{eq:SNR}.

\subsection{Theoretical template and halo bias fitting}
\label{sec 4.2 theory Cl}
\begin{figure}
    \centering
    \includegraphics[width=0.9\columnwidth]{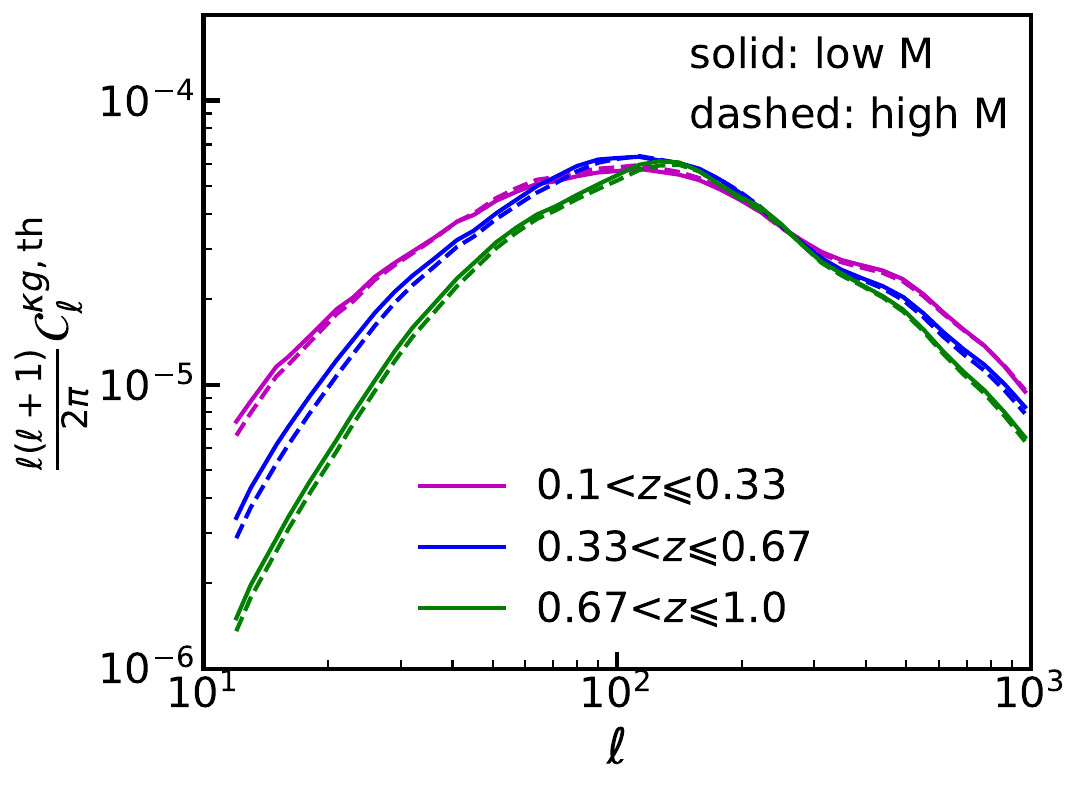}
    \caption{The theoretical cross power spectra for each sub-samples which set the bias equals to 1, calculated with \textsc{CCL}. Note that all theoretical calculations include Wiener filter and mask in the cross power spectra, with the former suppressing the power mainly at small scales, and the latter suppressing the power at all scales.}
    \label{fig:kg_b1.0}
\end{figure}

The theoretical template for $C^{\kappa g}$ is approximated as $b_{\rm g} C^{\kappa g,\rm th}$. $C^{\kappa g,\rm th}$ include the modification to the power spectrum caused by both the survey mask and the \emph{Planck} Wiener filter \footnote{The Wiener filter is $C_L/(C_L+N_L)$ where $N_L$ is the noise power spectrum and $C_L$ is the lensing power spectrum.}. The first step is to calculate the cross-power spectrum without the survey mask and the Wiener filter, using \myref{eqn:ckg} and setting the bias $b=1$. We use the Core Cosmology Library (CCL; \cite{pyccl}) for the calculations, with a fiducial \emph{Planck} 2018 cosmology \citep{Planck2018parameters}: $\Omega_m = 0.315, \Omega_\Lambda = 0.685, n_s = 0.965, h = H_0/(100 \ \rm km \ s^{-1} \ Mpc^{-1}) = 0.674$ and $\sigma_8 = 0.811$. We use HaloFit \citep{Smith2003, Takahashi2012, Mead2015} to calculate $P_{\rm mm}$. By default CCL uses \cite{Takahashi2012}. The galaxy redshift distribution $n(z)$ are calculated for each tomographic bin and mass bin, combining the photo-$z$ distribution and a Gaussian photo-$z$ error PDF with $\sigma_z=0.05$.\footnote{We numerically confirm that the adopted $\sigma_z$ in the range $[0.01,0.1]$ has negligible impact on the theoretical power spectra, as we may infer from the insensitivity of cross-power spectra to the redshift distribution (Fig.\,\ref{fig:kg_b1.0}). Therefore uncertainty in the $\sigma_z$ estimation has an insignificant impact on the current analysis. } The second step is to generate 100 full-sky maps of the power spectrum calculated in the first step. We then apply the Wiener filter and mask to these maps. We then obtain 100 cross-power spectra with the survey mask and Wiener. To correct for the pixelation effect of full-sky, a factor of $12 N_{\rm side}^2/4\pi$ is needed to obtain the correct power spectrum normalization. We use the average one as the theoretical template  $C^{\kappa g,\rm th}$. Fig.\,\ref{fig:kg_b1.0} shows the theoretical templates for the three redshift bins and two mass bins. The results of the two mass bins are slightly different due to the slightly different redshift distribution. 

The only free parameter in the theoretical model is the group density bias $b$. 
The $\chi^2$ for the fitting is 
\begin{equation}
    \chi^2 = \sum\limits_{\ell\ell^\prime}\Big[\big[C_{\ell}^{\kappa g,o}-bC_{\ell}^{\kappa g,{\rm th}}\big]\textbf{Cov}_{\ell\ell^\prime}^{-1}\big[C_{\ell^\prime}^{\kappa g,o}-bC_{\ell^\prime}^{\kappa g,{\rm th}}\big] \Big] \ .
\end{equation}
Here the sum is over the data points of each redshift/mass bin. Namely bias of each redshift/mass bin is independent of each other. The degree-of-freedom of the fitting is d.o.f $= N_D - N_f = N_D-1$, where $N_D$ is the number of data points, $N_f$ is the number of fitting parameter.
The best-fit value and the associated error are 
\begin{equation}\label{eq:b_fitting}
    b = \frac{\sum\limits_{\ell\ell^\prime} C_{\ell}^{\kappa g,\rm th}\textbf{Cov}_{\ell\ell^{\prime}}^{-1} C_{\ell^{\prime}}^{\kappa g,\rm o}}{\sum\limits_{\ell\ell^\prime} C_{\ell}^{\kappa g,\rm th}\textbf{Cov}_{\ell\ell^{\prime}}^{-1} C_{\ell^{\prime}}^{\kappa g,\rm th}} \ ,
\end{equation}
\begin{equation}\label{eq:b_error}
    \sigma_b= \Big[ \frac{1}{\sum\limits_{\ell\ell^\prime}C_\ell^{\kappa g,\rm th} \textbf{Cov}_{\ell\ell^\prime}^{-1} C_{\ell^\prime}^{\kappa g,\rm th}} \Big]^{1/2} \ .
\end{equation}

\subsection{The theoretically predicted halo bias}
The measured bias can be compared with the one from the theoretical prediction, given the estimated halo mass. We estimate the later by taking into account of the associated weighting in the Limber integral,
\begin{equation}\label{eq:b_model}
b^{\rm th} = \frac{\sum\limits_{i}^{N} b(M_i,z_i) W^{\kappa}(z_i) \chi(z_i)}{\sum\limits_{i}^{N} W^{\kappa}(z_i) \chi(z_i)} \ .
\end{equation}
Here $i$ denotes the $i$-th galaxy groups. $b(M,z)$ is the bias predicted by the theory, given the halo mass $M$ and $z$. We compare the measured bias with the prediction of three halo bias models \citep{Jing1998, SMT2001, Tinker2010}, calculated by \textsc{Colossus} \citep{colossus}. We define dark matter halos as having an overdensity of 180 times the background density of the universe. This mass definition is used for both \textsc{Colossus} and Y21 catalog.

\begin{figure}
    \centering
    \includegraphics[width=\columnwidth]{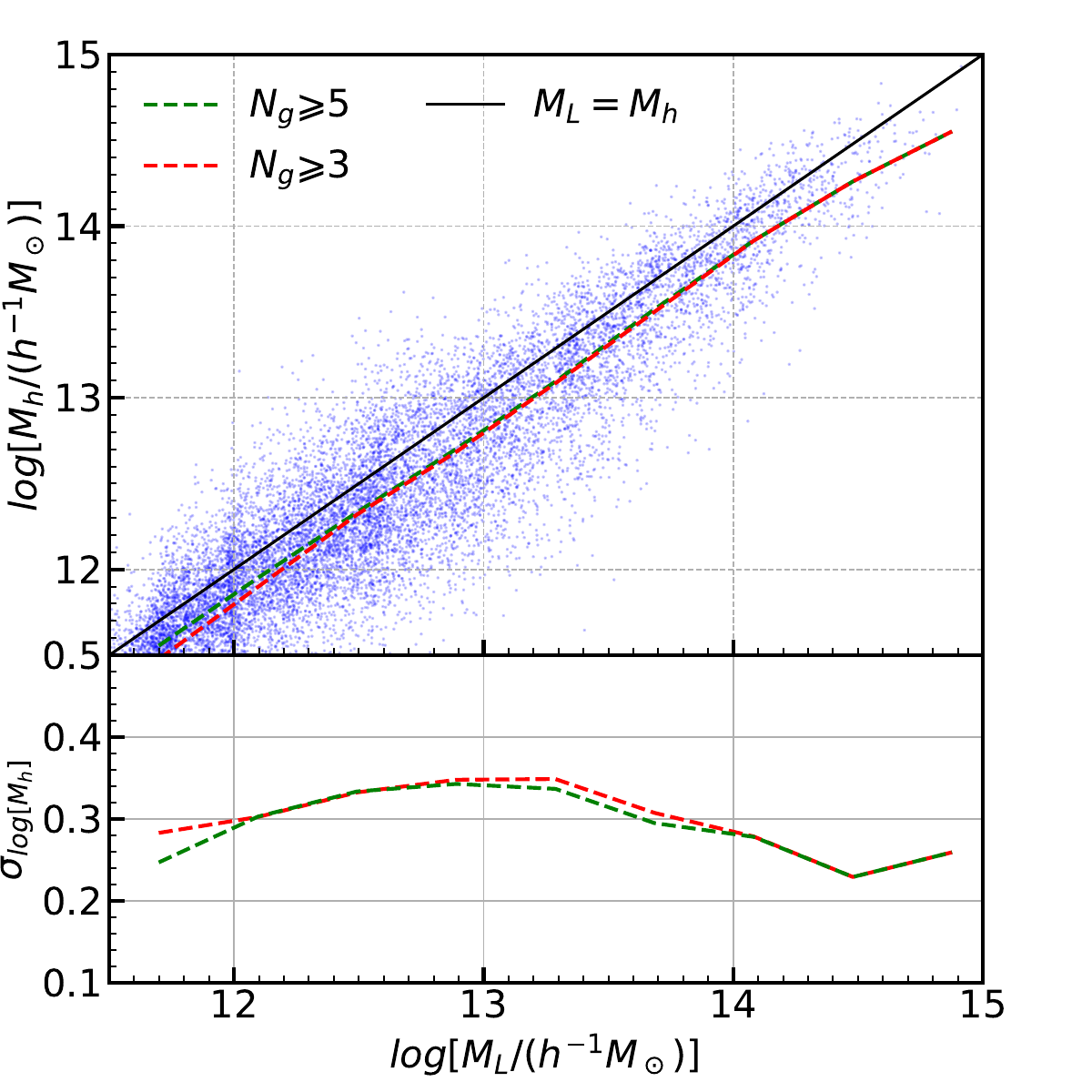}
    \caption{The comparisons between the assigned halo mass $M_L$ (based on the total group luminosity $L_{\rm tot}$) and the true halo mass $M_h$. The true halo masses $M_h$ are a small subset of groups in the mock group catalog constructed from MGRS. The dashed curves are shown for the average of true halo mass log$M_h$ deviates from the average of assigned halo mass in 10 log$M_L$ bins. The assigned halo mass is overestimated concerning the black $y=x$ solid line. The lower panels plot the scatter from the dashed lines, with red and green colors representing the richness cuts $N_g\geqslant3$ and $N_g\geqslant5$, respectively.}
    \label{fig:ML_Mh}
\end{figure}

We have to take care of the mass estimation uncertainty in estimating $b^{\rm th}$. Y21 uses the mock group catalog from the mock galaxy redshift survey (MGRS hereafter) to estimate the uncertainty, and finds an uncertainty of $\sim 0.2$ dex when $M\ge10^{13.5}$ and $\sim 0.45$ dex at the low mass end. Fig.\,\ref{fig:ML_Mh} we show the assigned halo mass $M_L$ (based on the total group luminosity $L_{\rm tot}$) versus the true halo mass $M_h$ in the B = 2.5 case, for $N_{\rm g}\geq 3(5)$. Besides a $\sim 0.3$ scatter in $\log M$, the averaged $\log M_L$ is also systematically higher than the true $\log M_h$. We model the $M_L$-$M_h$ relation with a log-normal PDF, with the systematic shift ($\langle \log M_h\rangle-\langle \log M_L\rangle$) and statistical scatter ($\sigma_{\log M}$) measured from Fig.\,\ref{fig:ML_Mh}. We replace $M_i$ in \myref{eq:b_model} as $M_h$ randomly drawn from this log-normal distribution, given $M_L$ and $z$. 

\section{Results}\label{section:results}
\begin{figure}
    \centering 
    \includegraphics[width=\columnwidth]{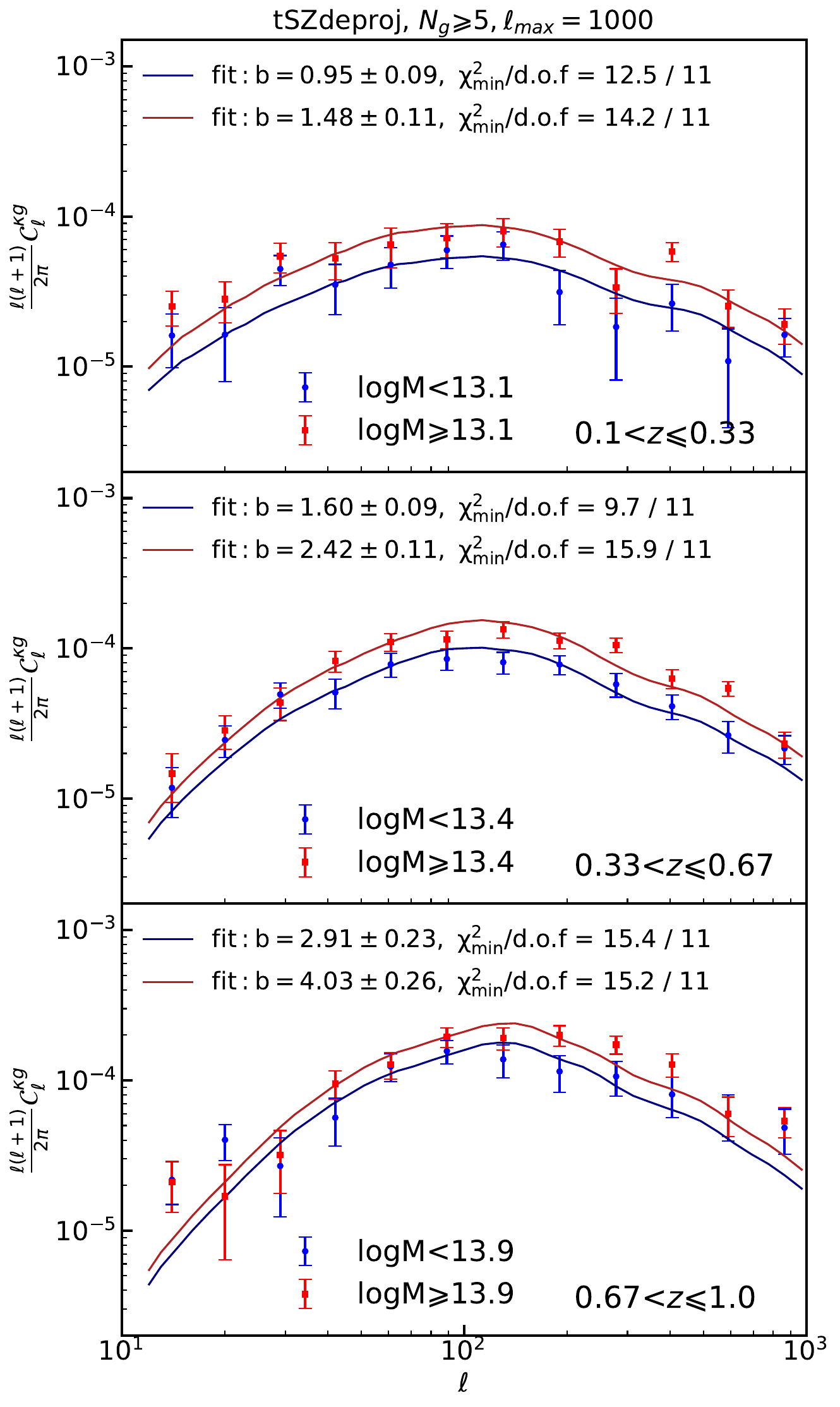}
    \caption{Measured cross power spectra \emph{Planck} CMB tSZ-deprojected lensing map and the baseline group sample with richness cut $N_{\rm g}\geq 5$ at 3 redshift bins and 2 mass bins. The data of each galaxy sub-sample is well fitted by the \emph{Planck} 2018 cosmology prediction with a scale-independent bias (solid curves). The bias is constrained with better than $10\%$ accuracy and shows clear mass and redshift dependence.  Note that the effect of the Wiener filter and survey mask is presented in both the measurement and in the theory curves. } 
    \label{fig:kg_tsz5}
\end{figure}

Fig.\,\ref{fig:kg_tsz5} shows the measured $C_\ell^{\kappa g,\rm o}$ with richness cut $N_g\geqslant5$, for three redshift bins at $0.1<z\leqslant1.0$ and two mass bins. We also show the best-fit theory curves, along with the best-fit bias and $\chi^2_{\rm min}$/d.o.f. Note that the effects of the Wiener filter and survey mask are presented in both the measurement and in the theory curves, instead of deconvolving the mask effect in the data part. We divide the $\ell$ bins from $\ell=10$ to 1000 in log space. We choose the small scale cut at $\ell$=1000 to avoid modeling small scale P(k). The large scale cut is to show the power of wide survey. The data and model are very consistent within $10 < \ell < 1000$ as shown in Fig.\,\ref{fig:kg_tsz5}.
\begin{figure}
    \centering   
    \includegraphics[width=0.9\columnwidth]{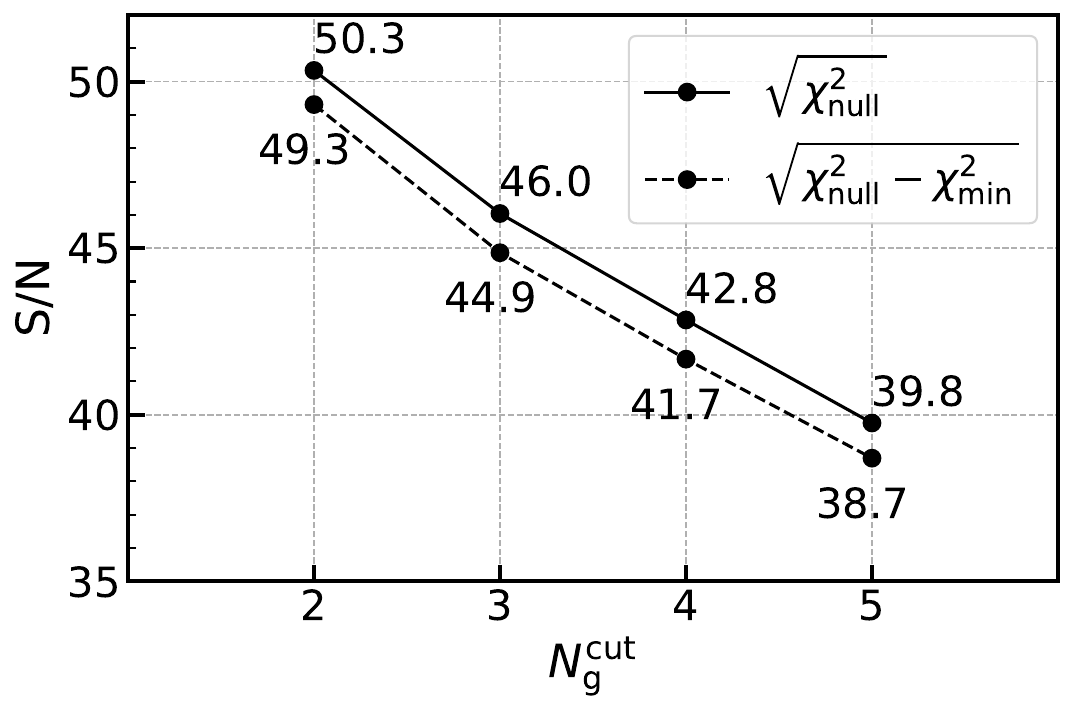}
    \caption{The total signal-to-noise ratio as a function of richness cut $N_{\rm g}$. The baseline group sample adopts a conservative cut with $N_{\rm g}=5$, but nevertheless achieves $38.7\sigma$ detection of CMB lensing-galaxy group cross-correlation. }
    \label{fig:SNR}
\end{figure}

\subsection{The detection significance}
The detections are significant at all six redshift-mass bins. We quantify the detection significance with two definitions of signal-to-noise ratio (S/N). One is data-driven, and describes the detection significance of a non-zero signal. This S/N $\equiv \sqrt{\chi^2_{\rm null}}$. Here 
\begin{equation}\label{eq:chi2_null}
    \chi^2_{\rm null} = \sum\limits_{\ell\ell^{\prime}} C_{\ell}^{\kappa g,\rm o} \textbf{C}_{\ell\ell^{\prime}}^{-1}  C_{\ell^{\prime}}^{\kappa g,\rm o} \ . 
\end{equation}
The six redshift-mass bins achieve $10$-$20\sigma$ detections of non-zero cross-correlation signal. 
Since the covariance between different redshift-mass bins is negligible, the total S/N combining all redshift/mass bins\footnote{This definition of S/N is identical to $\langle b\rangle/\sigma_{\langle b\rangle}$ for the optimally weighted $\langle b\rangle=\sum w_\alpha b_\alpha/\sum w_\alpha$.  Here $w_\alpha=b_\alpha/\sigma^2_{b_\alpha}$ is the weight to minimize fractional errors in $\langle b\rangle$.} is
 \begin{equation} \label{eq:SNR}
     \bigg(\frac{S}{N}\bigg)_{\rm total} = \sqrt{ \sum\limits_{\beta} \bigg(\frac{S}{N}\bigg)_{\beta}^{2} } \ .
 \end{equation}
Here $\beta=1,\cdots 6$ denotes the six redshift and mass bins. The total S/N for richness cut $N_{\rm g}\geq 5$ is $39.8$ for null discrimination and 38.7 for signal detection (Fig.\,\ref{fig:SNR}).

The other definition of S/N is fitting-driven, with S/N $\equiv \sqrt{\chi^2_{\rm null}-\chi^2_{\rm min}}$. Here 
\begin{equation}\label{eq:chi2_min}
    \chi^2_{\rm min} = \sum\limits_{\ell\ell^{\prime}} (C_{\ell}^{\kappa g,\rm o}-b^{\rm bestfit}C_{\ell}^{\kappa g,\rm th}) \textbf{C}_{\ell\ell^{\prime}}^{-1}  (C_{\ell^{\prime}}^{\kappa g,\rm o}-b^{\rm bestfit}C_{\ell^{\prime}}^{\kappa g,\rm th}) \ .
\end{equation}
The S/N defined in this way has the relation $\sqrt{\chi^2_{\rm null}-\chi^2_{\rm min}}$ = $b^{\rm bestfit}/\sigma_{b}$.\footnote{This relation holds when there is only one bias parameter. Thus it does not hold when combining measurements at all redshift and mass bins.} Therefore it represents the detection significance given the model. Meanwhile $\chi^2_{\rm min}$/d.o.f. justifies whether the model provides a good description of the data. Fig.\,\ref{fig:kg_tsz5} shows that the fitting with a single bias parameter returns reasonable $\chi^2_{\rm min}$/d.o.f. $\sim 1$. This means that scale dependence of the measured cross-power spectrum agrees with the \emph{Planck} 2018 cosmology prediction, and provides strong support that the measured signal indeed originates from CMB lensing-matter cross-correlation. Since the $\chi^2_{\rm min}$/d.o.f. $\sim 1$ and $\chi^2_{\rm min}\ll \chi^2_{\rm null}$,  $\sqrt{\chi^2_{\rm null}-\chi^2_{\rm min}}\simeq \sqrt{\chi^2_{\rm null}}$. Namely S/N defined in both ways are basically the same ($39.8$ versus $38.7$, for $N_{\rm g}\geq 5$ galaxy groups, Fig.\,\ref{fig:SNR}). 
The group bias is constrained with $5\%$-$10\%$ accuracy. It shows a significant increase with the halo mass and redshift, consistent with our theoretical expectation. And the p-value is 0.33 (0.22), 0.56 (0.14), 0.17 (0.17) for low-mass (high-mass) sample at $0.1<z\leq 0.33$, $0.33<z\leq 0.67$, $0.67<z\leq1$, respectively.

\begin{table}
	\centering
	\begin{tabular}{c|c|c|c|c|c|c|c} 
		\hline
		$z$ &  SMT2001 & Tinker2010 & Jing1998 \\
		\hline
		$0.1<z\leqslant0.33$ &  1.27/ 1.76 & 1.15 / 1.65 & 1.28 / 1.90\\
        $0.33<z\leqslant0.67$ & 1.68 / 2.47 & 1.55 / 2.41 & 1.78 / 2.79\\
        $0.67<z\leqslant1.0$ & 2.84 / 3.85 & 2.81 / 4.02 & 3.27 / 4.59\\
		\hline
	\end{tabular}
	\caption{Halo bias models by SMT2001, Tinker2010 and Jing1998. These three models are as a function of redshift and halo mass. The table shows $b(\rm M<M_{div})$ / $b(\rm M\geqslant M_{div})$ in three tomographic bins.}
	\label{tab:bias_model}
\end{table}
\begin{figure}
    \centering
    \includegraphics[width=0.9\columnwidth]{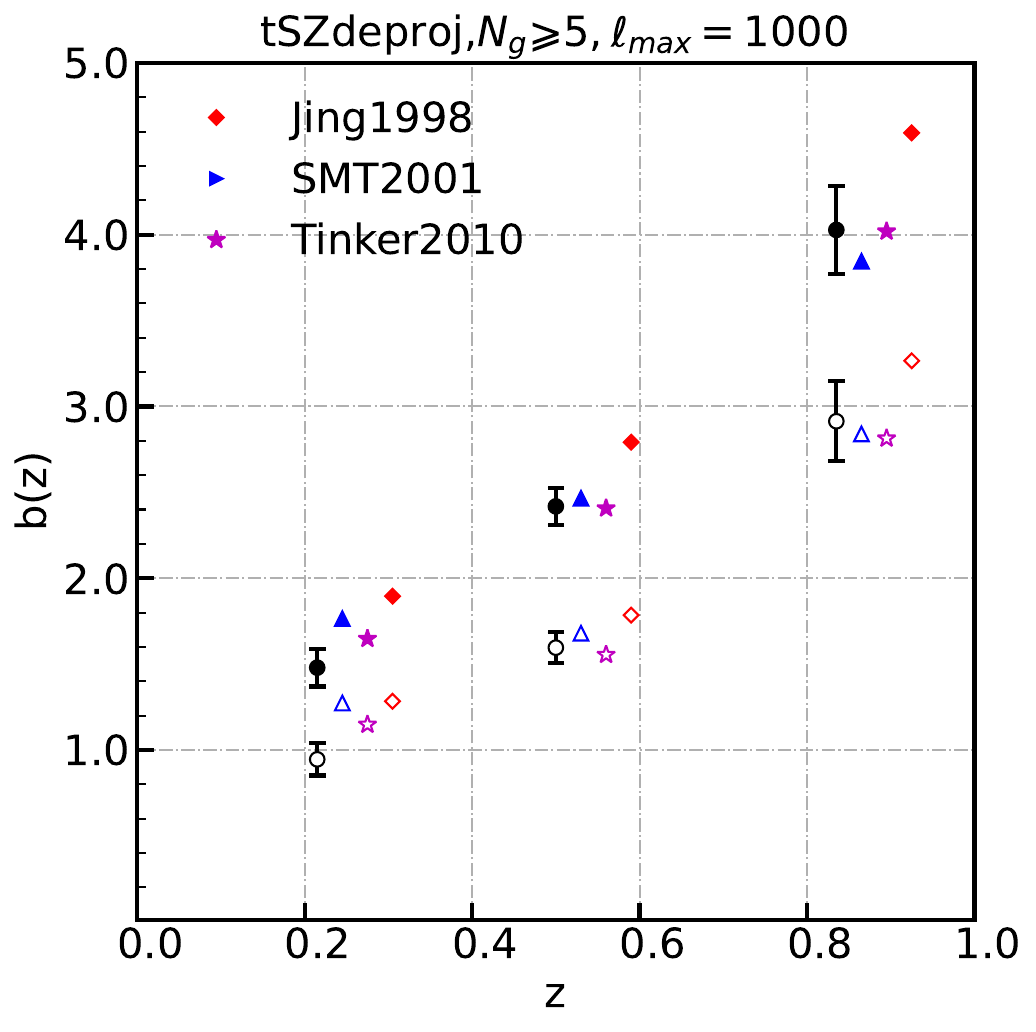}
    \caption{Comparisons between the fitting bias and three bias model predictions. High-mass (solid circle) and low-mass (empty circle) in black is our fitting bias for $N_g\geqslant5$ samples. The colored symbols are three bias model expectations, and they are shifted for better visualization.  Nearly all model predictions are within our fitting results, but there is a small shift in low-z samples.}
    \label{fig:b_tSZdeproj_Ng5}
\end{figure}

\subsection{Comparison with the theory prediction}
\label{subsec:comparison}
The predictions of three halo bias models are listed in Table.\,\ref{tab:bias_model}, while the comparison with the measurements are shown in Fig.\,\ref{fig:b_tSZdeproj_Ng5}.  The agreement is excellent at the two high redshift bins, providing strong support on the robustness of the measurement and demonstrating the usefulness of the group catalog. 

However, at the lowest redshift bin ($0.1<z\leq 0.33$), the measured halo biases are $\sim 23\%$ ($3.1\sigma$) lower than the theoretical expectations for the low-mass bin, and $\sim 16\%$ ($2.6\sigma$) lower for the high-mass bin. We have checked that this is not caused by statistical photo-$z$ errors or magnification bias. It may be caused by a systematic bias in group redshift estimation (e.g. due to the galaxy color-redshift degeneracy \citep{Zou2019}). If the true group redshift is higher, the resulting signal at the same large angular scale is smaller (Fig.\,\ref{fig:kg_b1.0}), leading to the wrong interpretation of a smaller $b_{\rm g}$. But the required bias in redshift is $\sim 0.1$ (Fig.\,\ref{fig:kg_b1.0}), likely too large to be realistic. 

We further discuss three remaining possibilities. (1) One is the misidentification of field galaxies of different redshifts as a group due to the line of sight projection effect. Such projection is most severer for low redshift group identification. The above two possibilities also affect the group-galaxy clustering, so it can be further tested by cross-correlating these galaxy groups with spectroscopic redshift galaxies. This is an issue for future investigation. (2) Another possibility is the existence of systematic overestimation of the halo mass. However, later tests with other group richness cut imply that it is unlikely the cause. Further complexity is that, for the low-mass bin, the low bias is caused by groups in the SGC (\S \ref{subsec:NGCSGC}). But for the high-mass bin, the low bias is caused by NGC groups. The discrepancy between NGC/SGC group implies that the low bias issue is related to the identification of galaxy groups under different observing conditions of NGC/SGC. This issue will be further discussed in \S \ref{subsec:NGCSGC}.  (3) The third possibility is that what we really measure is the amplitude $A$ of the cross-correlation compared to our fiducial theory.  Besides $b_{\rm g}$, other factors also affect A. For example, if the real $\sigma_8$ deviates from the fiducial value, or if the lensing reconstruction has certain bias such that $A_{\rm lens}\neq 1$, we have $A=b_{\rm g} (\sigma_8/\sigma_{8,{\rm fiducial}})^2 A_{\rm lens}$. Thus the low bias we find above may be caused by a $\sigma_8\sim 0.72$ or $A_{\rm lens}\sim 0.8$, or some combined effects. Interestingly, a recent work on a combined analysis of DESI galaxy clustering and galaxy-galaxy lensing also finds a low $\sigma_8\sim 0.65$ at the same redshift range (private discussion with Xu Haojie \footnote{Xu Haojie et al. in preparation}).

\begin{table}
    \centering
    \begin{tabular}{c|c|c|c|c|c}
        \hline
        z & $ f^{\rm threshold}$ & $b^{\rm bestfit} $ & $\chi^2_{\rm min}$ & \\
        \hline
        \multirow{3}{*}{$0.1<z\leqslant0.33$} & 0.3 & 0.95 $\pm$ 0.09(1.48 $\pm$ 0.11) & 12.3(13.7) \\ 
		~ & 0.5 & 0.95 $\pm$ 0.09(1.48 $\pm$ 0.11) & 12.5(14.2) \\
		~ & 0.7 & 0.95 $\pm$ 0.09(1.47 $\pm$ 0.11) & 11.8(14.7) \\
        \hline
        \multirow{3}{*}{$0.33<z\leqslant0.67$} & 0.3 & 1.60 $\pm$ 0.09(2.43 $\pm$ 0.11) & 9.3(15.9) \\ 
		~ & 0.5 & 1.60 $\pm$ 0.09(2.42 $\pm$ 0.11) & 9.7(15.9) \\
		~ & 0.7 & 1.58 $\pm$ 0.09(2.42 $\pm$ 0.11) & 9.2(16.4) \\ 
        \hline
        \multirow{3}{*}{$0.67<z\leqslant1.0$} & 0.3 & 2.94 $\pm$ 0.23(4.05 $\pm$ 0.26) & 15.9(15.0) \\ 
		~ & 0.5 & 2.91 $\pm$ 0.23(4.03 $\pm$ 0.26) & 15.4(15.2) \\
		~ & 0.7 & 2.92 $\pm$ 0.23(4.03 $\pm$ 0.25) & 15.4(14.5) \\
        \hline
    \end{tabular}
    \caption{The impact of $f^{\rm threshold}$ in defining the overdensity map on the analysis. Values in the parentheses are for the high-mass bins. The d.o.f. = 11. The agreement in the best-fit biases and $\chi^2_{\rm min}$ show that the fiducial $f^{\rm threshold}=0.5$ is valid for the analysis.  }    \label{tab:compare_threshold}
\end{table}

\subsection{Internal tests}
\label{sec5.3:internal}
\begin{figure}
    \centering
    \includegraphics[width=0.8\columnwidth]{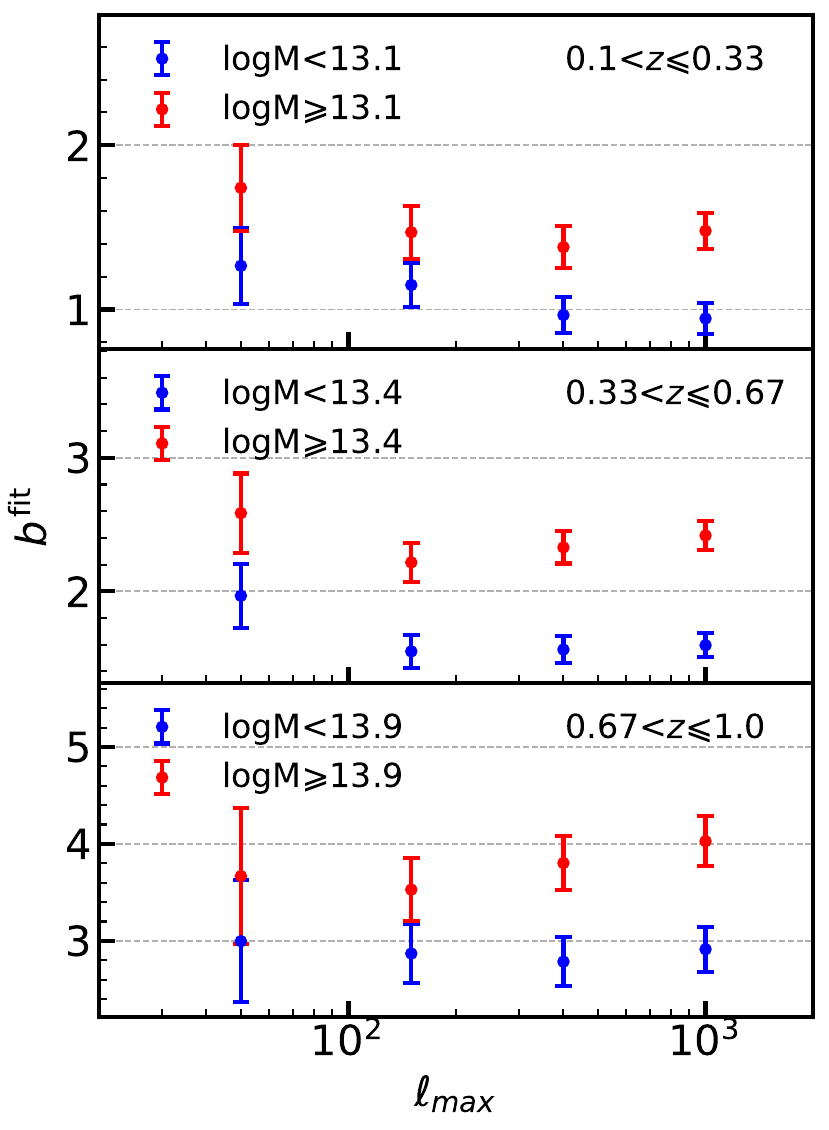}
    \caption{Consistency tests for $\ell_{\rm max}$ = 50, 150, 400, 1000, and $\ell_{\rm min}$ sets at 10. In each tomographic bin, fitting biases of low mass halos and the high mass halos are shown in blue circles and red circles, respectively.  }
    \label{fig:b_lmax}
\end{figure}

\begin{figure}
    \centering
    \includegraphics[width=0.8\columnwidth]{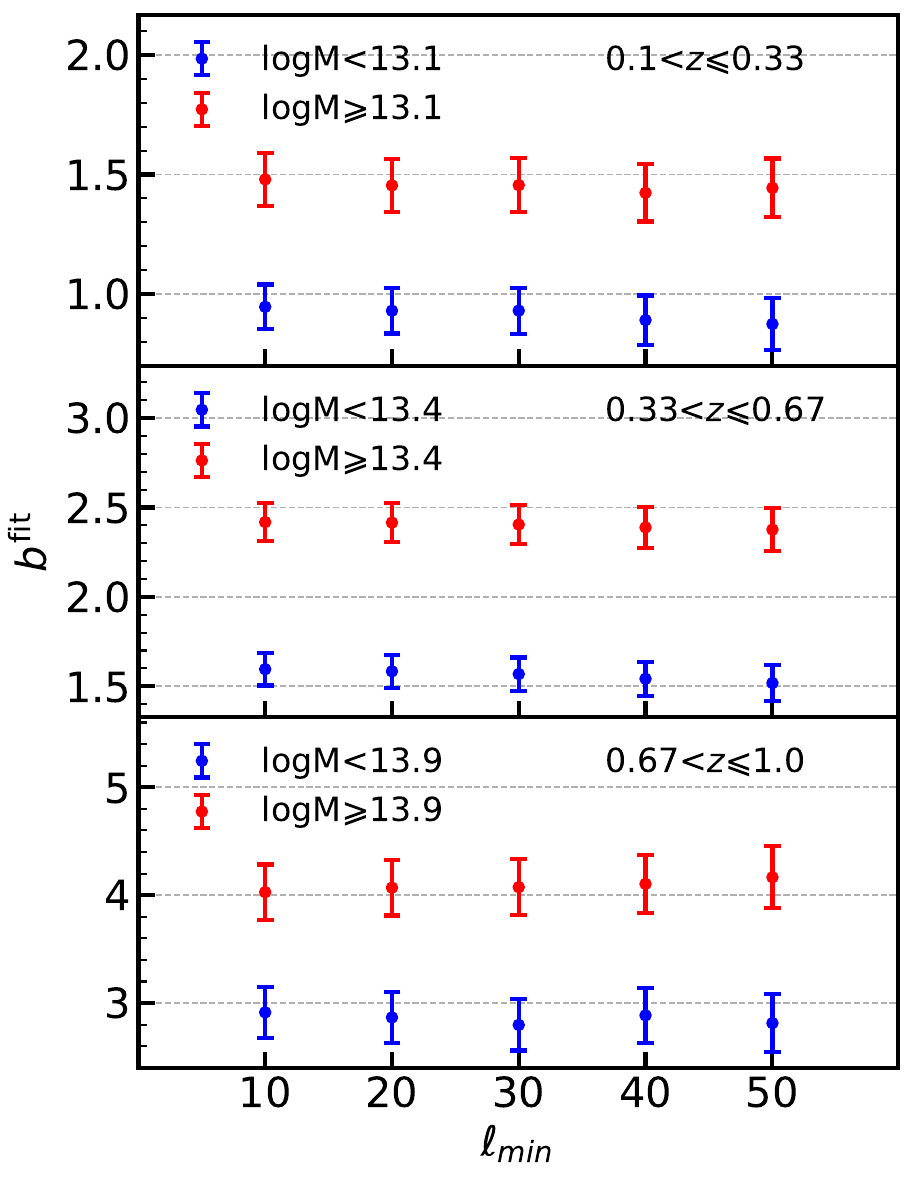}
    \caption{Consistency tests for $\ell_{\rm min}$ = 10, 20, 30, 40, 50, and $\ell_{\rm max}$ sets at 1000. In each tomographic bin, fitting biases of low mass halos and the high mass halos are shown in blue circles and red circles, respectively.}
    \label{fig:b_lmin}
\end{figure}

We test the impact of several parameters in the analysis. One is the threshold $f^{\rm threshold}$ in defining the group number overdensity map. Larger $f^{\rm threshold}$ leads to larger/more masks in the map. Table \ref{tab:compare_threshold} lists the best-fit bias results for $f^{\rm threshold}=0.3, 0.5, 0.7$. It shows that the impact of $f^{\rm threshold}$ is negligible, and the fitting result is stable around the fiducial $f^{\rm threshold}=0.5$. 

Another parameter is the adopted photo-$z$ scatter. The motivation of testing different photo-z scatter is, even though we don't have a true n(z) for the group catalog, the selection on the galaxy-richness of the groups can largely rule out the photo-z outliers. So we only need to show the photo-z scatter will not strongly bias our results. We test the cases of $\sigma_z=0.01, 0.05, 0.1$, and find that the impact on the best-fit bias can be ignored. $\sigma_z$ = 0.01 is the most optimistic value for galaxy groups photo-z estimation, and 0.1 is the worst. We choose the $\sigma_z$ = 0.05 as the fiducial one. The fractional error caused by the $\sigma_z$ is on the theoretical templates, and also corresponds to the error on the estimated bias $b_g$. Under different choices of $\sigma_z$, the biggest impact on the theoretical template is for the lowest redshift bin, the fractional error is $\sim$ 3.5\% for $\sigma_z$ = 0.1, and $\sim$ 1.7\% for $\sigma_z$ = 0.01. This is insignificant compared to the statistical error which is about 10\% at the lowest redshift. For the other two redshift bins, the fractional errors are less than 1\%. 

Another parameter is the $\ell$ range ($\ell_{\rm min}<\ell<\ell_{\rm max}$) for the fitting. These $\ell$ cuts are for general considerations of mitigating potential complexities at small scales (e.g. baryonic physics) and large scales (e.g. imaging systematics \citep{Rezaie2020}). Furthermore, in practice the S/N is negligible  at $\ell<10$  due to large cosmic variance, and is also negligible at  $\ell>1000$ due to overwhelming reconstruction noise in \emph{Planck} lensing map. Therefore we adopt the fiducial $(\ell_{\rm min},\ell_{\rm max})=(10,1000)$.  Fig.\,\ref{fig:b_lmax} shows the consistency tests by varying $\ell_{\rm max}=50,150,400,1000$, while fixing $\ell_{\rm min}=10$. The best-fit $b$ is in general insensitive to $\ell_{\rm max}$. But we notice a trend of decreasing $b$ with increasing $\ell_{\rm max}$, for the two low redshift and low-mass bins. If we only use the large scale data, the discrepancy between the measured bias and the theoretical prediction at low redshift may vanish. However, since the errorbars increase with aggressive $\ell_{\rm max}$ cut, the deviation between $b$ of different $\ell_{\rm max}$ cut is only $\sim 1\sigma$. Fig.\,\ref{fig:b_lmin} shows the consistency tests for $\ell_{\rm min}$ = 10, 20, 30, 40, 50, while fixing $\ell_{\rm max}$ = 1000. The best-fit $b$ has negligible dependence on $\ell_{\rm min}$. Therefore we demonstrate the robustness of our analysis against the choice of $(\ell_{\rm min},\ell_{\rm max})$.

\subsection{Richness dependence}
We check the richness cut $N_{\rm g}$ adopted for this work. $N_{\rm g}$ has two-fold impacts on our results. First, lower $N_{\rm g}$ results in more groups, therefore higher measurement S/N. The S/N increases from S/N = 40 of $N_{\rm g}=5$ to S/N = 50 of $N_{\rm g}=2$ (Fig.\,\ref{fig:SNR}). But on the other hand, lower $N_{\rm g}$ leads to lower purity in the group sample. This does not decrease the S/N, since even field galaxies contribute to the cross-correlation signal. However, it complicates the comparison with theory. This is the reason that we choose $N_{\rm g}=5$ as the fiducial richness cut. Nevertheless, since the average richness is a good proxy of halo mass, we expect the measured bias to increase with the richness. Therefore, despite the purity issue, measuring the dependence of cross-correlation on the richness cut is still a useful internal check of the measurement. 

\begin{figure}
    \centering
    \includegraphics[width=0.8\columnwidth]{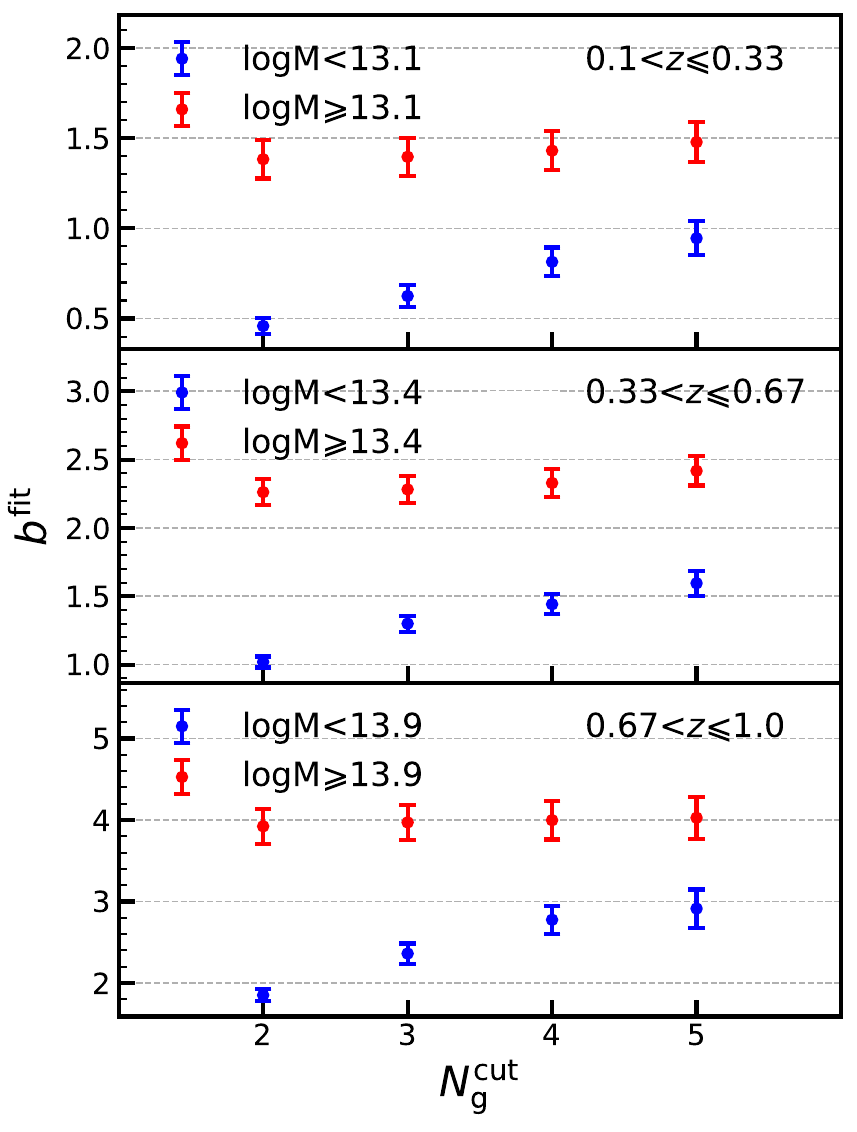}
    \caption{The dependence of best-fit bias on the richness cut $N_g\geqslant$ 2, 3, 4, 5. For the higher mass bins (red), the results are consistent with different richness cuts, demonstrating the $N_g\geqslant5$ groups are reliable enough. The changes in the lower mass bins (blue) are due to the fact richness is a mass proxy, so that it introduces an extra selection on the sample mass.}
    \label{fig:b_Nmember}
\end{figure}

The results for $N_{\rm g}=2,3,4,5$ are shown in Fig.\,\ref{fig:b_Nmember}.  For high mass bins, the best-fit halo biases are insensitive to $N_{\rm g}$. This result is expected, since the measured group bias is determined by the majority of halos whose masses are above but close to the mass cut. But for low mass bins,  $b_{\rm g}$ shows a significant increase with increasing $N_{\rm g}$, for all three redshifts. The majority of halos now have richness $N_{\rm g}$ and the mean halo mass/bias then decreases with $N_{\rm g}$.

The $N_{\rm g}\geq 2$ group sample reveals an interesting problem.  The low redshift and low mass bin have the best-fit bias $b_{\rm g}=0.46\pm 0.04$. This extremely low bias can not be explained within the standard $\Lambda$CDM cosmology, since the halo bias has a lower bound around $0.7$ (e.g. \citet{Jing1998}). It can not be explained by an overall shift  in the photo-$z$, since the required shift is too large to be realistic. A more plausible cause is the misidentification of field galaxies in overlapping lines of sight as groups. In this case, the contamination to ``low'' redshift groups is most likely from high redshift field galaxies, leading to weaker clustering at large angular scales (Fig.\,\ref{fig:kg_b1.0}) and underestimation of group bias. This possibility should be further checked with future DESI spectroscopic redshift data, which will enable the measurement between these groups with galaxy distribution in true redshift. 

\subsection{Detection of the significant impact of tSZ}
The thermal Sunyaev Zel'dovich (tSZ) effect can in principle contaminate the reconstructed lensing maps and bias the lensing-galaxy cross-correlation measurement. Previous work such as \citep{Baxter2018, MartinWhite_LRGs, Dong2021} do not detect a significant impact of tSZ. \cite{Omori2019a} constructed a lensing systematic map through the tSZ map to better quantify its impact. They found that its impact on SPT+Planck CMB lensing-galaxy cross-correlation is significant at $\la 10^{'}$, but insignificant at $\ga 20^{'}$. 

\begin{figure}
    \centering
    \includegraphics[width=0.9\columnwidth]{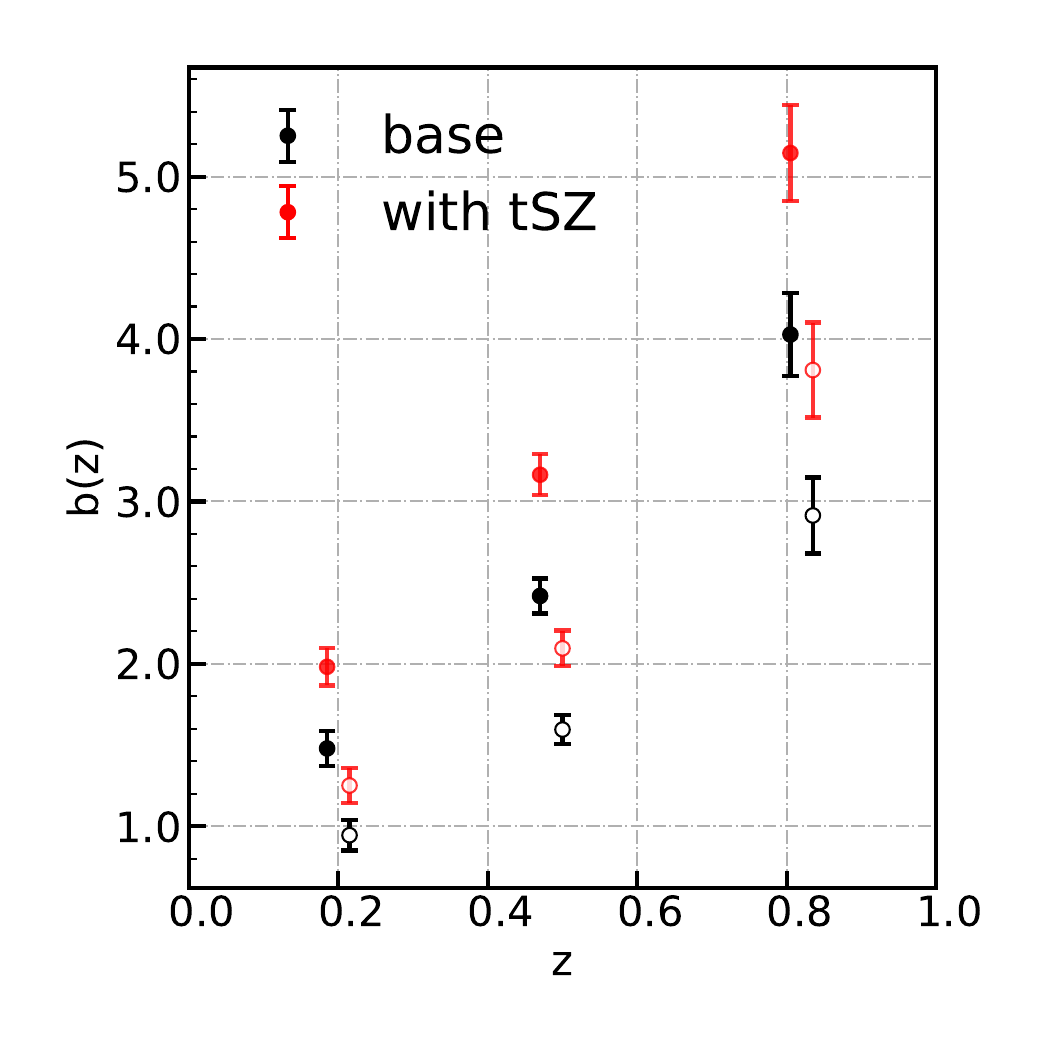}
    \caption{The comparisons of group biases measured with or without tSZ-deprojected. High-mass (filled circle) is shifted relative to low-mass (open circle) for better visualization. The impact of tSZ is significant at all redshift/mass bins. This is the reason that we choose the tSZ-deprojected lensing map as the baseline data. }
    \label{fig:b_basetsz}
\end{figure}

To check the impact of tSZ, we also measure the cross-correlation of groups with the \emph{Planck} minimum-variance (MV) lensing map reconstructed from the SMICA CMB map. The S/N is 44.5 for null discrimination and $43.4$ for signal detection. However, we find significant difference to the baseline case (with the lensing map reconstructed from tSZ-deprojected CMB temperature map). In terms of the group biases, they are $\sim30\%$ higher at all redshift and mass bins, and the differences are all significant ($\sim 5\sigma$, Fig.\,\ref{fig:b_basetsz}). This finding is interesting. The major difference to previous works is that we use galaxy groups instead of galaxies. These groups have mean mass $\sim 10^{13} M_\odot/h$, and we have detected the tSZ effect with high significance \citep{Chen2021}. The tSZ effect associated with these galaxy groups is likely responsible for our finding \footnote{The contamination to the lensing map is quadratic in tSZ, $\kappa(\vec{\ell})\propto \sum_{\vec{\ell}_1} f(\vec{\ell}_1,\vec{\ell}_2) y(\vec{\ell}_1) y(\vec{\ell}_2)$ \citep{Hu2002}.  Here $y$ is the tSZ y-parameter and $\ell_2=\vec{\ell}-\vec{\ell}_1$. This biases the group-CMB lensing cross-correlation by an additive term  $\propto \sum_{\vec{\ell}_1} f(\vec{\ell}_1,\vec{\ell}_2) \langle y(\vec{\ell}_1) y(\vec{\ell}_2) \delta^*_{\rm g}(\vec{\ell})\rangle$. Therefore group/cluster scale tSZ effect ($\ell_1\gtrsim 1000$) can propagate into degree scale cross-correlation measured in this work.  Furthermore, since the same groups/clusters in defining $\delta_{\rm g}$ also produce $y$, one-halo term  contributes to $\langle y(\vec{\ell}_1) y(\vec{\ell}-\vec{\ell}_1) \delta^*_{\rm g}(\vec{\ell})\rangle$. In contrast, this term vanishes in the cross-correlation with field galaxies since they do not occupy the same halos. This difference may be responsible for the different tSZ impacts on the galaxy-CMB lensing and group-CMB lensing. }.  

This tSZ contamination is by itself useful for understanding gastrophysics. But the full investigation of its impact requires both measuring the tSZ effect of these groups with the \emph{Planck} y-map, and propagating it into the \emph{Planck} lensing map. These are beyond the scope of this work. Therefore for the purpose of this paper, we choose to use the tSZ-deprojected lensing map as our baseline analysis. To do so, we sacrifice the detection significance ($45\sigma$ to $39\sigma$), but gain cross-correlation measurement free of tSZ-related bias.

\section{Conclusions and discussion}\label{section:conclusions}
In this work, we present the measurements of the DESI groups/clusters $\times$ \emph{Planck} CMB lensing cross-correlations with tomographic analysis. We achieve an overall S/N $\simeq 40$ for groups of richness $N_g\geq 5$. This allows us to measure the group density bias and its redshift/mass dependence to $\sim 10\%$ accuracy. Besides routine consistent tests of the measurement, the value-added group catalog allows comparing the measured bias with the theory prediction. The agreement is excellent for $z>0.33$ groups, but with a $\sim 3\sigma$ discrepancy for $z<0.33$ groups. 

The above measurement of group bias is mainly for the validation of cross-correlation measurement. We have not attempted to constrain cosmology with the above measurements. One reason is due to uncertainties in the true redshift distribution of galaxy groups. The situation can be significantly improved by DESI. DESI will directly measure the redshifts of some group member galaxies. For the rest, through cross-correlation with DESI galaxies, the group redshift distribution can be tightly constrained. DESI redshift information will also improve uncertainties in estimating the group purity and completeness. Another issue is the impact of tSZ contamination on the lensing map. We find a significant difference whether we use the tSZ-deprojected map. For the same groups, we have detected a strong tSZ signal by stacking groups with the Planck $y$-map. We will further estimate whether the directly detected tSZ is consistent with the found impact on group-CMB lensing cross-correlation. With these extra pieces of information, we expect the usefulness of group-CMB lensing cross-correlation in constraining cosmology.

\section*{Acknowledgements}
This work made use of the Gravity Supercomputer at the Department of Astronomy, Shanghai Jiao Tong University. This work is supported by the National Key R\&D Program of China (2020YFC2201602, 2018YFA0404504, 2018YFA0404601, 2020YFC2201600), National Science Foundation of China (11621303, 11653003, 11773021, 11773048, 11890691), the 111 project No. B20019, the CAS Interdisciplinary Innovation Team (JCTD-2019-05), and the China Manned Space Project (NO. CMS-CSST-2021-A02, CMS-CSST-2021-B01). JY acknowledges the support from China Postdoctoral Science Foundation (2021T140451). FYD is supported by a KIAS Individual Grant PG079001 at Korea Institute for Advanced Study.

\section*{Data Availability}
No new data were generated or analyzed in support of this research. The data underlying this article will be shared on reasonable request to the corresponding author.

\bibliographystyle{mnras}
\bibliography{example} 

\appendix
\counterwithin{figure}{section}

\section{Magnification bias}

Magnification bias modulates the number density of galaxies. Lensing magnifies the surface area and therefore decreases the number density by $1/\mu\sim 1-2\kappa$. But it also amplifies the galaxy flux and changes the galaxy number density above a flux limit or within a flux bin. Overall the observed galaxy number density $\delta_g^{\rm obs}$ is changed from the intrinsic overdensity $\delta_g^{\rm int}$ to
\begin{equation}
    \delta_g^{\rm obs} = \delta_g^{\rm int} + g_\mu\kappa \ ,
\end{equation}
Here $g_\mu$ depends on the galaxy flux/magnitude \citep{Yang2017,MartinWhite_LRGs,Wietersheim-Kramsta2021}. It then biases the group-convergence angular power spectrum, 
\begin{equation}\label{eq:kmu}
    C_\ell^{\kappa g} \rightarrow C_\ell^{\kappa g} + C_\ell^{\kappa \mu} \ .
\end{equation}
Here $C_\ell^{\kappa\mu}$ denotes the convergence-magnification power spectra.

\begin{figure}
    \centering
    \includegraphics[width=0.8\columnwidth]{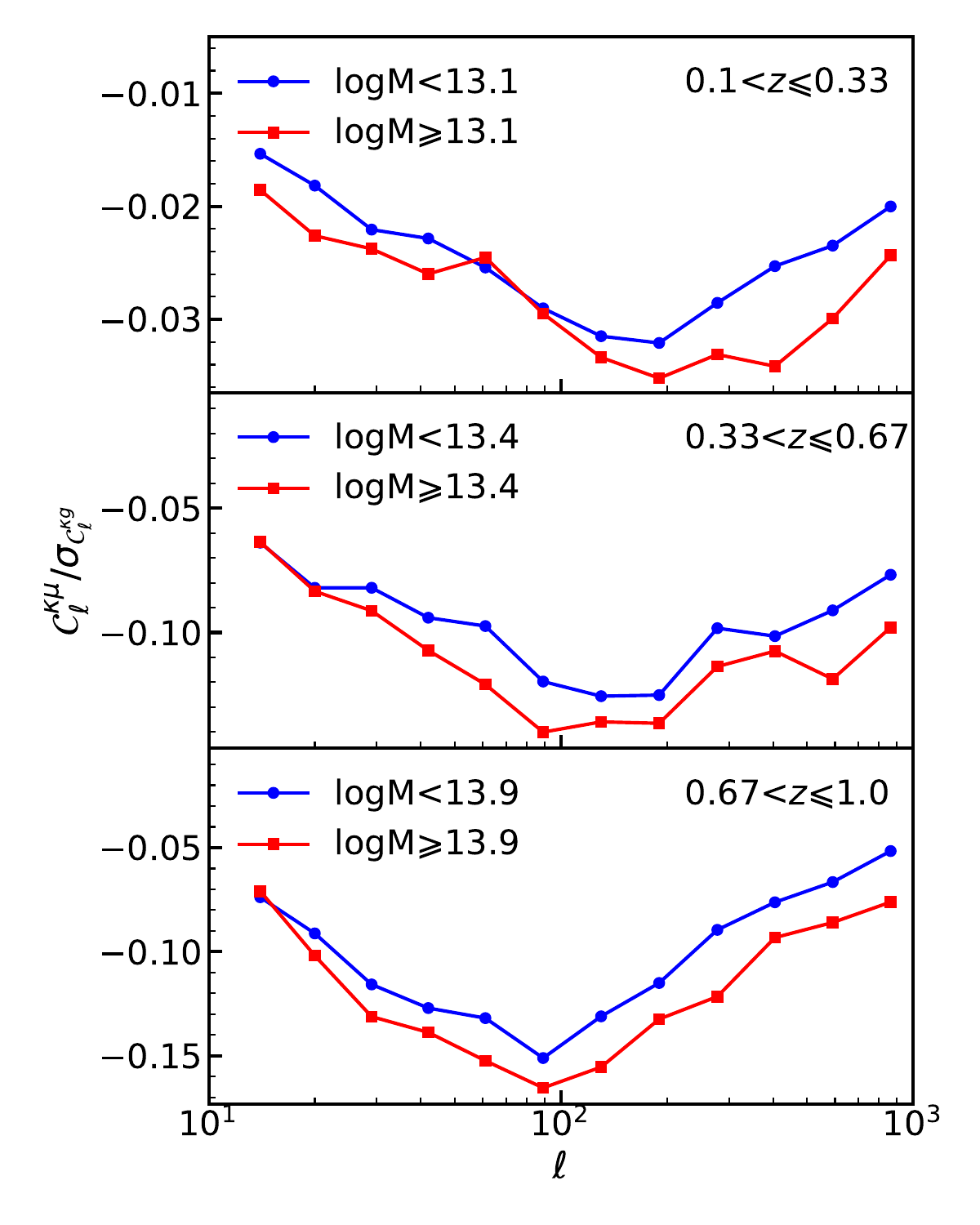}
    \caption{The magnification bias term of \myref{eq:kmu} as a fraction of the uncertainty of observed spectra. Here, we choose g = -2 for group samples, so that the $C_{\ell}^{\kappa \mu}$ signal is positive to the observed cross-correlation signals. }
    \label{fig:kmu}
\end{figure}

The contamination ($C^{\kappa \mu}/C^{\kappa g}$) increases with redshift, but remains $<1\%$ (Fig.\,\ref{fig:kmu}). The induced systematic shift in the group bias $b_{\rm g}$ is 
\begin{equation}
    \delta b_{\rm g}=\frac{\sum\limits_{\ell\ell^\prime} C_{\ell}^{\kappa g,\rm th}\textbf{C}_{\ell\ell^{\prime}}^{-1} C_{\ell^{\prime}}^{\kappa \mu}}{\sum\limits_{\ell\ell^\prime} C_{\ell}^{\kappa g,\rm th}\textbf{C}_{\ell\ell^{\prime}}^{-1} C_{\ell^{\prime}}^{\kappa g,\rm th}}\ .
\end{equation}

We find $\delta b_{\rm g}$ = -0.0040 (0.0051), -0.0149 (0.0197), -0.0430 (0.0519) from low to high redshift bins. The values in the parentheses are for the high-mass bins. Here we adopt $g_\mu=-2$, since the selection criteria of groups have a weak dependence on the galaxy flux and the area magnification should be dominant in its magnification bias. Compared with the statistical error $\sigma_b$, the systematic shift caused by magnification bias is $\leq 0.2\sigma_b$. Therefore it is negligible.

\section{NGC/SGC}
\label{subsec:NGCSGC}
Consistency tests have also been performed for the north galactic cap (NGC) and south galactic cap (SGC) \citep{Zhao2021, Zarrouk2021}, to test the concordance of different observational conditions, sample selections, or potential breakdown of cosmological isotropy.

\begin{figure}
    \centering
    \includegraphics[width=0.8\columnwidth]{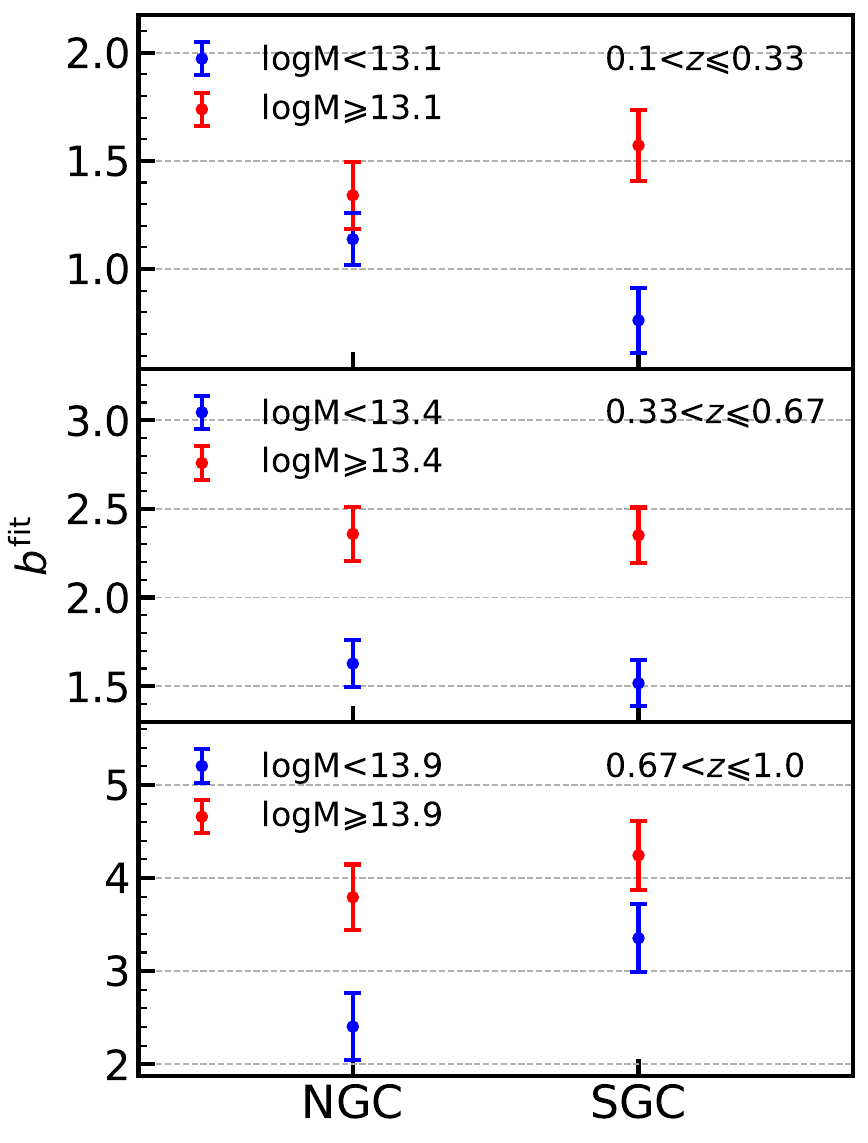}
    \caption{Comparisons between NGC and SGC for $N_g\geqslant5$ samples. We find that some are not consistent within $1\sigma$ error, that could come from the variety of mass and redshift distribution of NGC and SGC, we show the mass and redshift distributions in Fig.\,\ref{fig:Number_NSGC_z1_less}. }
    \label{fig:b_NGCSGC}
\end{figure}
\begin{figure}
    \centering
    \includegraphics[width=0.8\columnwidth]{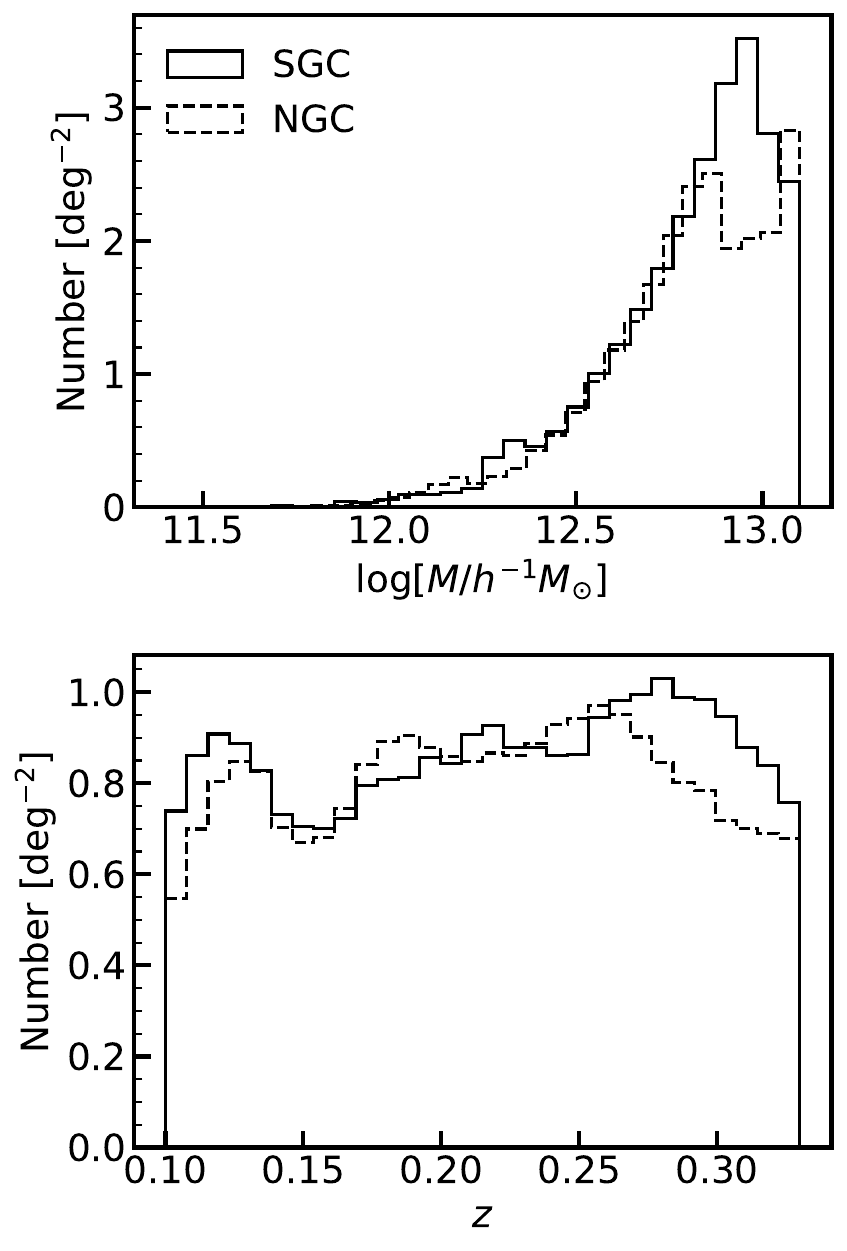}
    \caption{The mass and redshift distribution of NGC and SGC samples in the $0.1< z\leqslant 0.33$ and $\log M\leqslant 13.1$ bin. The shown differences may result into the different bias shown in Fig.\,\ref{fig:b_NGCSGC}. }
    \label{fig:Number_NSGC_z1_less}
\end{figure}

Fig.\,\ref{fig:b_NGCSGC} shows the best-fit biases for NGC and SGC respectively. The $0.33<z\leq 0.67$ redshift bin results are consistent. But for the $0.1<z\leq 0.33$ redshift bin and low-mass bin, the SGC bias is lower than the NGC bias by $\sim 1\sigma$. Although it is statistically insignificant, it is likely the cause of the discrepancy between the theory and the measurement discussed in \S \ref{subsec:comparison}. However, for the high-mass bin, the SGC bias is higher than the NGC bias, although the difference is within $1\sigma$. A similar discrepancy is also found for the low-mass bin at $0.67<z\leq 1$. Detailed investigation of these discrepancies is beyond the scope of this paper, but they may be related to the different observational conditions which affect the group identification. The mass and redshift distributions are indeed different between the NGC and SGC samples (Fig.\,\ref{fig:Number_NSGC_z1_less}). The mean number density/mass/redshift are 120/deg$^2$, $3.65\times 10^{13}M_\odot$, 0.38 for NGC, and 134/deg$^2$, $3.63\times 10^{13}M_\odot$, 0.39 for SGC. These differences result in $0.82\%$ difference in the  theoretically predicted bias ($\langle b\rangle =$ 2.42 for NGC and 2.44 for SGC).

We show the mass and redshift distributions of low-mass NGC and SGC groups in $0.1<z\leqslant0.33$. The amount of groups in NGC and SGC is 1089259 and 947505, respectively. However, it is yet to be investigated in the future whether this discrepancy is due to cosmic inhomogeneity or observational systematicity.

\begin{figure}
    \centering
    \includegraphics[width=\columnwidth]{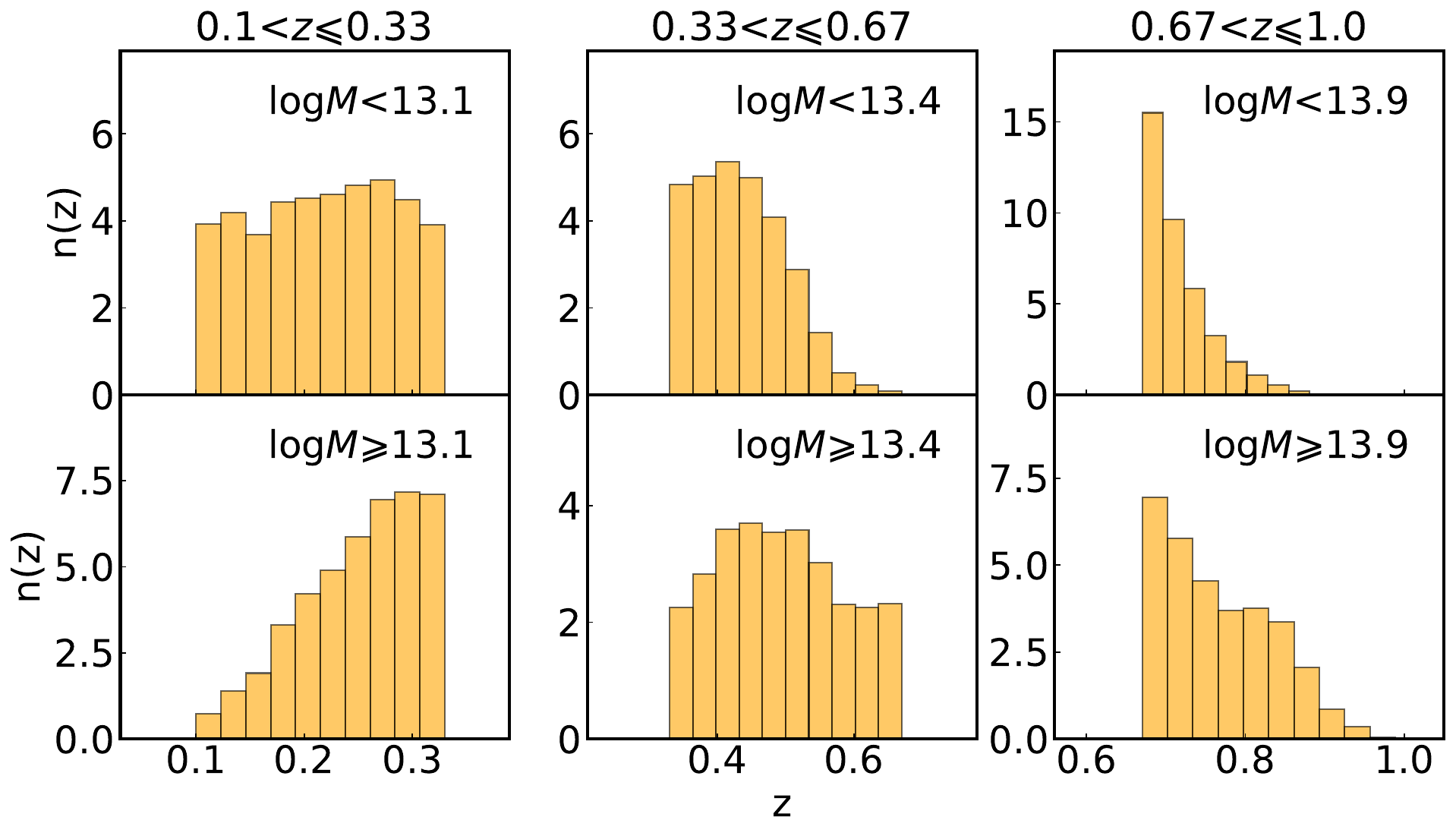}
    \caption{The redshift distribution for three redshift bins and two mass bins. Within each redshift bin, the redshifts of groups for small masses are lower. This is consistent with our experience, more massive galaxy groups are observed at high redshifts.}
    \label{fig:B3}
\end{figure}
In Fig.\,\ref{fig:Number_NSGC_z1_less} \ref{fig:B3}, we compare the mass distribution and redshift distribution (in the unit of square degree) between the NGC and SGC in the first redshift bin and the low-mass sub-sample. We find that the group distribution is inhomogeneous and the number of groups in NGC is greater than in SGC. We also find that the distribution of group mass is complementary at $M\sim10^{13}$ between NGC and SGC. The difference of redshift and mass distribution could cause a big difference of bias in this sub-sample.

% \begin{figure}
%     \centering
%     \includegraphics[width=0.8\columnwidth]{figures/figB4.pdf}
%     \caption{Comparisons fitting bias of NGC and SGC for $N_g\geqslant5$ samples with three bias model predictions labeled same as Fig.\,\ref{fig:b_tSZdeproj_Ng5}. We find that some fitting biases are not consistent with bias models within $1\sigma$ error. But the bias model predictions between NGC and SGC do not change a lot. }
%     \label{fig:B4}
% \end{figure}

%\section{Impact of $\sigma_{z}$ on the theoretical power spectrum}
%Given the 
%As mentioned in \S \ref{sec5.3:internal}, we show the different choices of $\sigma_z=0.01, 0.05, 0.1$ in %Fig.\,\ref{fig:D}. We find the changes in the theoretical cross power spectra is very small, so that the %impact on the best-fit bias is negligible.

%\begin{figure}
%    \centering
%    \includegraphics[width=\columnwidth]{figures/figD.pdf}
%    \caption{The impact of $\sigma_z$ on the theoretical cross power spectra. The impact is negligible %for the current analysis. }
%    \label{fig:D}
%\end{figure}

\section{Uncertainties in S/N estimation}
\begin{figure}
    \centering
    \includegraphics[width=0.8\columnwidth]{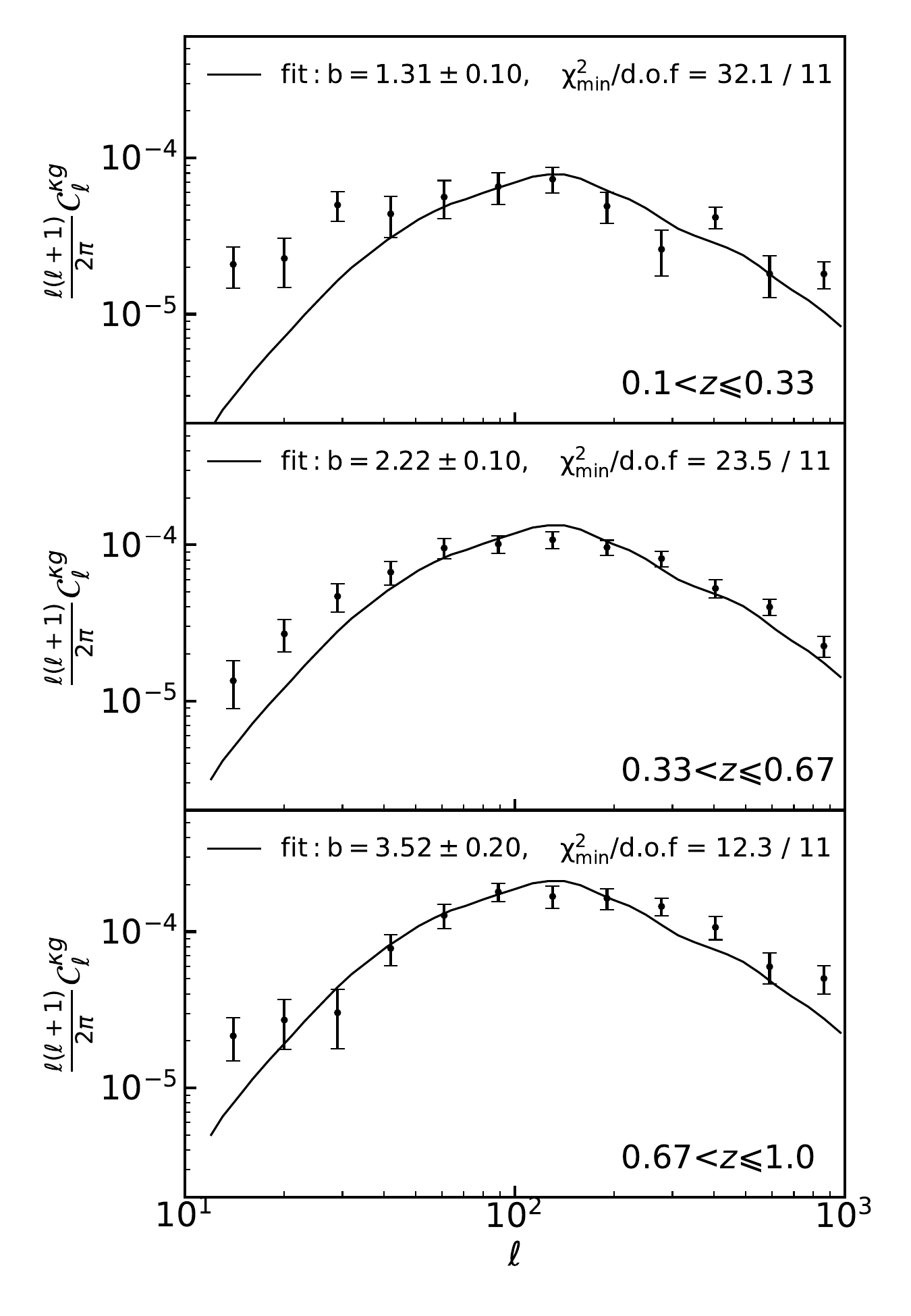}
    \caption{Similar to Fig.\,\ref{fig:kg_tsz5}, but merging the low-mass and high-mass bins. The merged group sample produces S/N = $32.1$ for null discrimination and $31.0$ for signal detection, slightly lower than the combined S/N of low and high-mass bins. We further discuss uncertainties of estimating S/N in the main text. }
    \label{fig:E}
\end{figure}

When combining measurements at different redshift and mass bins into the total S/N, we have approximated that the measurement errors are uncorrelated between bins. This certainly applies to different redshift bins. However, for the low-mass and high-mass samples of the same redshift, it is valid only where their shot noises dominate over their intrinsic clusterings. Numerically we find that shot noise dominates at $\ell\ga 100$. So summing over $\rm (S/N)^2$ to obtain the total S/N is valid. Nevertheless, at $\ell\la 100$, the cross-correlation measurement errors between low and high-mass bins are indeed correlated. A simple sum over $\rm (S/N)^2$ could lead to overestimation of the total S/N.  

To test the robustness of the estimated S/N, we merge the two mass bins within each redshift bin, and measure the cross power spectra (Fig.\,\ref{fig:E}). S/N of the merged group sample is $ \sqrt{\chi^2_{\rm null}} = 32.1$ and $\sqrt{\chi^2_{\rm null} - \chi^2_{\rm min}} = 31.0$. This is slightly lower than S/N = $39.8 (38.7)$ by the estimation with $\rm \sqrt{(S/N)^2_{\rm low}+(S/N)^2_{\rm high}}$. The reduction in S/N may be caused by correlated measurement error between the low and high-mass bins. However, this is not the only cause of S/N reduction. When we merge two groups, we lose the extra information on the mass and the mass-dependent clustering, so the S/N of the merged group sample is expected to be lower. Fig.\,\ref{fig:kg_tsz5} shows that the low-mass and high-mass cross power spectra may have different shapes. This may be responsible for the larger $\chi^2_{\rm min}$/d.o.f. and lower S/N of the merged sample. Nevertheless, since there is no significant difference between the two estimations of S/N, we leave the investigation of their origin into future works. 

%\begin{figure}
%    \centering
%    \includegraphics[width=0.8\columnwidth]{figures/figC.pdf}
%    \caption{The theoretical matter power spectra $C_{\rm mm}$ for low-mass (solid curves) and %high-mass (dashed curves) sub-samples, and the shot noise curves = $4\pi f_{\rm sky}/N_{g}$ (dotted %lines). Because the low and high-mass bins almost have the same $N_{g}$, we plot one line for each %redshift bin. Shot noise dominates over intrinsic clustering at $\ell\ga 100$. This implies that if we use %galaxies as tracers, the cross-correlation S/N can be further improved, as in \citet{Dong2021}. }
%    \label{fig:C}
%\end{figure}

\bsp	% typesetting comment
\label{lastpage}
\end{document}